\documentclass[11pt]{article}
\usepackage{jheppub}
\usepackage{amssymb}

\newcommand{\be}{\begin{equation}}
\newcommand{\ee}{\end{equation}}
\newcommand{\beqa}{\begin{eqnarray}}
\newcommand{\eeqa}{\end{eqnarray}}

\def\gke{G_{_{\!K\! E}}}
\def\Gtt{G_{tt}}
\def\Grr{G_{rr}}
\def\Gxx{G_{xx}}
 
 \def\dd{{\mathrm{d}}}

\def\le{\left}
\def\ri{\right}
\def\cN{{\cal N}}
\def\cO{{\cal O}}
\def\cG{{\cal G}}
\def\parl{\parallel}
\def\imm{{\rm Im}~}
\def\non{\nonumber}

\def\s{\sigma}

\def\a{\alpha}
\def\b{\beta}
\def\g{\gamma}
\def\k{\kappa}

\def\del{\delta}
\def\e{\epsilon}
\def\o{\omega}
\def\z{\zeta}

\title{Probes on D3-D7 Quark-Gluon Plasmas}

\author[a]{\'Alvaro Maga\~na,}
\author[a]{Javier Mas,}
\author[a]{Liuba Mazzanti,}
\author[b]{ Javier Tarr\'\i o.} 

\affiliation[a]{Departamento de F\'\i sica de Part\'\i culas and
Universidade de Santiago de Compostela,
E-15782, Santiago de Compostela, Spain.}
\affiliation[b]{Institute for Theoretical Physics and Spinoza Institute, Universiteit Utrecht, 3584 CE, Utrecht, The Netherlands.}

\emailAdd{alvaromagana@gmail.com} 
\emailAdd{javier.mas@usc.es} 
\emailAdd{mazzanti@fpaxp1.usc.es} 
\emailAdd{l.j.tarriobarreiro@uu.nl}

\abstract{We study the holographic dual model of quenched flavors immersed in a quark-gluon plasma with massless dynamical quarks in the Veneziano
limit. This is modeled by embedding a probe D7 brane in a background where the backreaction of massless D7 branes has been taken into account. The
background, and hence the effects, are perturbative in the Veneziano parameter $N_f /N_c$, therefore giving small shifts of all magnitudes like the
constituent mass, the quark condensate, and several transport coefficients. We provide qualitative results for the effect of flavor degrees of freedom
on the probes. For example, the meson melting temperature is enhanced, while the screening length is diminished. The drag force is also enhanced.}

\keywords{AdS/CFT correspondence, Energy loss in plasmas, Flavor physics} 

\subheader{ITF-UU-12/19, SPIN-12/17}

\begin{document}
\maketitle


\section{Introduction and results}

Since the times of Rutherford's experiment, the most direct way of obtaining valuable information about the microscopic structure of a chunk of matter consists in sending hard probes that penetrate the system and, thereafter, analyzing the outcome.
Nowadays, the quark-gluon plasma remains a challenging system for which a vast amount of data keeps accumulating along the last five years. The peculiarity of the heavy ion experiments performed at RHIC and CERN is that the probes are created spontaneously inside the plasma, and not introduced by hand by the experimenters. 

From these data, we have been able to see that many static and dynamical properties of probe quarks and mesons are dramatically affected when immersed in a
high density and temperature medium.
Examples of phenomenologically relevant properties include the screening of the quark-antiquark potential, the melting of mesons (but not of glueballs) at a certain temperature, or 
 the dramatic jet suppression effects observed on back to back hard probes. It makes sense to ask how much of these effects can be ascribed to the
quarks that live in the fundamental representation, and how much to the gluons, in the adjoint.

The quark-gluon plasma should be described by a ground state of a theory which contains dynamical quarks and gluons. At least for temperatures at or
above the deconfinement transition $T\sim 170$ MeV, the quarks $u$ and $d$ can be considered to be massless, and therefore fully dynamical. The
strange quark, $s$, with mass $\sim 10^2$ MeV should also be considered dynamical, albeit not massless in the same context. For the $c$ and $b$ quarks
the assumption that they are quenched is justified by their heaviness, and the probes discussed in the present analysis may be interpreted as
describing the phenomenology of these quarks. The $t$ quark is too massive to be included in the regime of validity of our model, as discussed in the
text (actually, the $b$ quark is on the edge of the regime of validity).

The problem of the propagation of probes inside a plasma is addressed in this paper by using the AdS/CFT correspondence. Our analysis is a
continuation of the work initiated in \cite{D3D7QGP}. The holographic setup consists of a solution to type IIB supergravity in the presence of D7
sources which are dual to massless matter transforming in the fundamental representation (quarks). It can be seen as a deformation of $\cN=4$ SYM
where $N_f$ D7 branes are introduced in the Veneziano limit: $N_f\to\infty$ with $N_f/N_c$ fixed. Using the presence of a large number of sources we
smear them along their 2-dimensional transverse space, recovering some isometries of the internal space present in the unflavored type IIB solution.
The geometry is dual to a field theory in which the beta function is positive. Therefore there is a Landau pole where a UV completion is needed. As
we will review below, this is captured by the supergravity solution, and limits the current approach to phenomenology in the far IR limit of the
theory. On top of this background we will consider a finite number $N_f'\ll N_f,N_c$ of massive probe D7s. We are interested in the dynamics of these
$N_f'$ flavors, since they will give information about the behavior of massive quenched quarks in a plasma of dynamical massless quarks and gluons.

The paper is structured as follows. In section \ref{sec:setup} we shall review the background and its properties. We will also choose a criterium for
comparing two theories, one without and one with dynamical flavor. The aim is to discriminate between replacing degrees of freedom (from adjoint to
fundamental) from merely piling them up. In section \ref{sec:probes} we will discuss the embedding of D7 probes in this background. The effect of the
unquenched massless flavors is to induce logarithmic corrections to the profile which, in principle, should affect the expression for the $\langle
\bar \psi \psi \rangle$ quark condensate. Deprived of a direct dictionary, in section \ref{sec:masscond} we will show how to extract the correct value
of the condensate from the renormalized boundary action. Furthermore, by studying the condensate, we observed that the phase transition between a
confined and a deconfined flavor (meson melting point) occurs at lower values of the quotient $M_q/T$. Finally, we also argue that the constituent mass
can be computed reliably for masses not very large as compared to the temperature, and see how the presence of a sea of quarks increases it. 

In section \ref{sec:conduc} we address the problem of the conductivity of a $U(1)_B$ current on the D7 probe. In order to do this, we shall
accommodate for an additional non-vanishing value of the $A_0$ component of the world-volume gauge field. This study suggests that the fundamental
(quarkonic) degrees of freedom present in the background have larger cross section than the adjoint (gluonic) degrees of freedom. This point is supported also by the study of the drag force of heavy quarks. Furthermore, the pair production of quarks is shown to contribute less to the polarization of the vacuum than the pair production of gluons.

Section \ref{sec:qpotential} is devoted to the quark-antiquark potential. This is performed in the standard way by looking at the spatial Wilson line for long times, and confirms, as was found in similar cases, that the dynamical flavors tend to screen the color charge.

Finally, in sections \ref{sec:dragforce} and \ref{sec:jetquenching} we study the problem of energy loss of probes in the plasma, due to the drag force and the jet quenching effect respectively. In both cases the presence of fundamental matter enhances the loss of energy, thus suppressing back-to-back jets and stopping heavy quarks moving through the plasma. In this last case the heavy quark, modeled holographically by a string hanging from a brane near the boundary, feels an effective temperature which is larger in the presence of unquenched flavor with respect to the unflavored case, and this leads to a larger absorption of momentum by the presence of a larger black hole in the worldsheet of the string.

\section{The flavored ${\cal N}=4$ SYM gravitational dual} \label{sec:setup}

The gravitational background we will consider was proposed in \cite{D3D7QGP} as a holographic dual to a quark-gluon plasma with unquenched, massless flavors in the Veneziano limit $N_c\to \infty$ and $N_f/N_c$ finite, with $N_c$ ($N_f$) the number of color (flavor) degrees of freedom. 

This scenario corresponds to a solution of type IIB supergravity in 10 dimensions with $N_f$ D7 flavor branes. Those branes act as sources in the
supergravity action for the $F_1$ RR field strength, and backreact into the geometry created by a stack of $N_c$ coincident D3 branes at the tip of a
Calabi-Yau (CY) cone. The flavor branes are smeared along their transverse directions to recover (some of) the isometries of the Sasaki-Einstein (SE)
manifold that defines the base of the CY cone.
The construction is very general and can accommodate any 5 dimensional SE manifold $X_5$, for which the metric can be written as a $U(1)$ fiber bundle over a 4 dimensional K\"ahler-Einstein (KE) space
\be
ds^2_{X_5} = ds^2_{K\! E} + (d\tau + A_{K\! E})^2\ , \label{desecinco}
\ee
where $A_{K\!E}$ is the K\"ahler-Einstein potential, \emph{i.e.}, $dA_{K\!E} = 2J$ with $J$ the K\"ahler-Einstein form. Despite its generality, it is useful to keep two 
main examples in mind. The first one, $X_5=S^5$, is the usual original standard example of holographic dual to $\cN=4$ supersymmetric gauge theory,
where the base KE space is $CP^2$. The second is the singular conifold $X_5=T^{1,1}$, which is dual to the Klebanov-Witten quiver, and the KE base is
$S^2\times S^2$, in this case.

The 10d metric ansatz introduced in \cite{D3D7QGP} is given by the following line element
 \beqa
ds_{10}^2 &=& G_{tt} dt^2 + G_{xx} dx_3^2 + G_{rr} d r^2
+ \gke ds_{ K\! E}^2 + G_{\tau\tau} (d\tau + A_{K\! E})^2 \ .
\label{deformedads5bh}
\eeqa

The $F_1$ RR field strength sourced by the smeared D7 brane squashes the fiber relative to the base in (\ref{desecinco}). Hence the difference
$G_{\tau\tau}-\gke$ is a measure of the backreaction of the flavor branes onto the geometry, which will be parameterized by a dimensionless parameter
$\epsilon_h$. The configuration breaks the supersymmetries from ${\cal N}=4$ to ${\cal N}=1$. Furthermore, finite temperature will break all
supersymmetries. To keep the discussion short we will specialize to the case of $AdS_5\times S_5$.

 The solution to the sourced IIB system depends on a dimensionful (energy) scale\footnote{Actually, the solution depends also on a UV scale, $r_*$,
acting as a Wilsonian renormalization scale, from which a completion of the theory is needed. We will comment later on this.}, $r_h$, and a
dimensionless parameter, $\epsilon_h$. The scale $r_h$ is the horizon location and sets the temperature scale $T= r_h/\pi R^2+ {\cal O}(\epsilon_h)$, with $R$ the curvature of the manifold. The second parameter, $\epsilon_h$, is proportional to the number density of flavor branes
\be
\epsilon_h = \frac{\lambda_h}{8\pi^2} \frac{N_f}{N_c} \ ,
\ee 
where $\lambda_h=4\pi g_s e^{\Phi_h} N_c\gg1$ is the ``running" 't Hooft parameter evaluated at the IR scale $r_h$. 
 
In the case of the five sphere, the SE manifold corresponds to $CP^2$, so the KE metric is given by
$
ds^2_{S^5} = ds^2_{CP^2} + (d\tau + A)^2 
$, where\footnote{ The ranges of the angles in $CP^2$ and the fiber are $0\leq\chi<\pi~,~~0 \leq \theta\leq \pi~,~~0 \leq \varphi< 2\pi~,~0\leq \xi< 4\pi$ and $0\leq \tau< 2\pi $.}
\beqa
ds^2_{CP^2} &=& \frac{1}{4} d\chi^2 + \frac{1}{4}\cos^2\frac{\chi}{2} (d\theta^2 + \sin^2\theta d\varphi^2) + \frac{1}{16}\sin^2 \chi(d\xi + \cos\theta d\varphi)^2 \ ,\label{intmetr}\\
A_{K\!E} &=& \frac{1}{2} \cos^2 \frac{\chi}{2} (d\xi + \cos\theta d\varphi) \ .
\eeqa
For the ansatz \eqref{deformedads5bh}, we have the perturbative solution \cite{D3D7QGP}
\beqa
G_{tt} &=& - \frac{r^2}{R^2} \left( 1 -\frac{r^4}{r_h^4}\right) \ , \qquad \qquad
G_{xx} =\frac{r^2}{R^2} \ , \label{solttxx}\\
G_{rr} &=& \frac{R^2}{r^2} \left( 1 -\frac{r^4}{r_h^4}\right)^{-1}\left( 1 + \frac{\epsilon_h}{4} + \epsilon_h^2 \left( \frac{11}{24}- \log \frac{r_h}{r}\right) + {\cal O}(\epsilon_h^3) \right) \ , \\
\gke &=& R^2\left( 1 + \frac{\epsilon_h}{12} + \epsilon_h^2\left( \frac{5}{288} - \frac{1}{12}\log\frac{r_h}{r}\right) + {\cal O}(\epsilon_h^3) \right) \ , \\
G_{\tau\!\tau} &=& R^2\left( 1 - \frac{\epsilon_h}{12} + \epsilon_h^2\left( \frac{1}{32} + \frac{1}{8}\log\frac{r_h}{r}\right) + {\cal O}(\epsilon_h^3) \right) \label{bulkmettau} \ , \\
\phi &=& \epsilon_h \log\left(\frac{r}{r_h}\right) + \epsilon_h^2
\left( \frac{1}{6}\left(1 + 3\log \frac{r}{r_h}\right)\log \frac{r}{r_h} + \frac{1}{16} \log_2\left(1 - \frac{r_h^4}{r^4}\right) \right)+ {\cal O}(\epsilon_h^3)
 \label{solphi}. 
\eeqa
Here we absorbed the constant $\Phi_h$ in the definition of the dilaton field $\phi=\Phi-\Phi_h$.
The $F_1$ RR field strength has the form $F_1=Q_f (d\tau +A_{KE})$, with $Q_f$ a constant proportional to the number of ``flavor" D7 branes, $N_f$.
For the self-dual RR field strength one has the usual ansatz $F_5=Q_c (1+\star) \varepsilon_5$ with $Q_c$ a constant proportional to the number of
``color" D3 branes, $N_c$, and $\varepsilon_5$ the volume form of the SE manifold $X_5$.
 
The above solution should be understood as an effective IR one in the Wilsonian sense. Namely, the perturbative solution that the reader can find in
\cite{D3D7QGP} contains another scale, $r_*$, beyond which the solution itself should be replaced by the correct UV completion ($r_*$ would be the analog of
the Z boson mass $M_Z$ for the Fermi theory of electro-weak interactions). One can show that the dilaton blows up at a radius $r_{LP} \sim
r_*e^{1/\epsilon_h}$. From here we see that, besides the possibility of obtaining a perturbative solution, another reason to keep $\epsilon_h\ll1$ is
to have the scale of the Landau pole exponentially separated from the Wilsonian cutoff scale.  The bad UV behavior of the solution is understandable
on physical grounds, and signals the fact that the beta function is positive on the QFT side. Hence, the physical coupling constant is known to
diverge at a certain UV (Landau) pole. In the spirit of effective theories, we expect that IR observables are mildly affected by whatever UV
completion is added. In fact, as explained in \cite{D3D7QGP}, all corrections to physical results computed at a scale $r<r_*$ are modulated by factors
${\cal O}(r/r_*)$. Therefore, if we insist in investigating the far IR limit of the theory, $r\sim r_h$, we can discard the corrections coming from
the UV completion. This has been realized formally, in the solution (\ref{solttxx})-(\ref{solphi}), by sending $r_*\to \infty$. The
thermodynamic quantities that were evaluated in \cite{D3D7QGP} are an example of IR quantities that do not depend on the UV completion. The
temperature of this background is obtained from the absence of conical singularities after Wick-rotating the time direction by
\be
T = \frac{r_h}{\pi R^2}\left( 1- \frac{\epsilon_h}{8}-\frac{13}{384} \epsilon_h^2 + \cdots \right) \ . \label{bulktemp}
\ee
Also the energy density can be adequately described within the present framework, giving 
\be
\varepsilon = \frac{3\pi^2}{8} N_c^2 T^4 \left( 1 + \frac{1}{2}\epsilon_h + \frac{1}{3} \epsilon_h^2 + \cdots \right) \ .\label{bulkendens}
\ee

The present paper deals with comparisons of quantities as computed in two different theories, one with flavor and the other one without flavor. Still there are
several parameters that can be chosen to be equal in both theories. In \cite{D3D7QGP,D3D7QGPb} an extensive discussion has been provided on the
different possible choices. 
Here we will stick to one choice, by keeping the temperature $T$ and the energy density $\varepsilon$ fixed. From (\ref{bulktemp}) and (\ref{bulkendens}) this simply amounts to shifting the value of 
$r_h$ and $N_c$ suitably.
\beqa
r_h & = & r_h^{(0)}\left ( 1 + \frac{\epsilon_h}{8} + \frac{19}{384} \epsilon_h^2 + \cdots\right) \label{rhshift} \ ,\\
N_c &=& N_c^{(0)} \left( 1 -\frac{\epsilon_h}{4} - \frac{7}{96} \epsilon_h^2 + \cdots \right) \label{Ncshift} \ ,
\eeqa
with 
\be
r_h^{(0)} = \pi R T \ , \qquad N_c^{(0)} = \frac{2}{\pi T^2}\sqrt{\frac{2\varepsilon}{3}} \label{zeroconst} \ ,
\ee
the values in the flavorless background.
In this way, rather than merely piling up degrees of freedom, we are partially replacing adjoints by fundamentals while keeping a measure of the total sum of them
unaltered\footnote{Alternatively, the entropy density could have been chosen but results remain unchanged up to first order in $\epsilon_h$, as $s$
and $\varepsilon$ only differ at order $\epsilon_h^2$}. For example, we will also need to shift the 't Hooft coupling 
\be
\lambda_h = 4\pi g_s e^{\phi_h} N_c(\epsilon_h)=\lambda_h^{(0)} \left( 1 -\frac{\epsilon_h}{4} - \frac{7}{96} \epsilon_h^2 + \cdots \right) \ .
\label{thooftsh}
\ee

The fact that the solution \eqref{solttxx}-\eqref{solphi} is perturbative in $\epsilon_h$ adds another source of divergence in the large $r$ limit,
coming from the truncation of the series at a finite order.
Terms of the form $\epsilon_h \log(r/r_h)$ stop being perturbative corrections unless
 $r\ll r_\Lambda\equiv r_h e^{1/\epsilon_h}$ ($\sim 10^2 r_h$ for $\epsilon_h= 0.24$ , see section 6 of \cite{D3D7QGP}). In particular this implies we should not consider
placing arbitrarily massive probe quarks in this background, as the point $r_{min}$ of maximum approach of the D7 should be way below $r_{min} \ll r_\Lambda$ for
the computations to be reliable.

\section{D7 probes on the flavored background}\label{sec:probes}

In this section, and due to computational limitations, we will restrict ourselves to ther first order in $\epsilon_h$ for the backreacted solution. We
will consider a class of embeddings for the probe D7 branes that, apart from extending along the Minkowski 4-space, wrap a non-compact cycle in the
cone
\be
\tau = \tau_0 \ , \qquad \chi = \chi(r) \label{embedd} \ .
\ee
For example, in the supersymmetric case, the embeddings are given by 
$
r_0 = r \sin \left( \chi(r)/2 \right)
$,
where the mass of the probe quark is proportional to $r_0$
and, hence, massless embeddings have $\chi = 0$.
 For a general profile, $\chi(r)$, the induced metric will read from (\ref{intmetr}) 
\beqa
ds^2_8 &=& G_{tt} dt^2 + G_{xx}d\vec x^2 + 
\left( G_{rr}+ \frac{\gke}{4}\chi'^2\right) dr^2 + \frac{\gke}{4} \cos^2\frac{\chi}{2} \left( d\xi^2 + d\theta^2 + d\phi^2 + 2 \cos \theta d\xi d\phi \right) \nonumber \\
&& + \frac{G_{\tau\tau}-\gke}{4}\cos^4\frac{\chi}{2}(d\xi + \cos \theta d\phi)^2 \ .\label{indmetr}
\eeqa
Notice that, when the backreaction vanishes, $G_{\tau\tau}-\gke=0$, the probe brane wraps an $S^3\subset S^5$, but this is no longer the case when
$\epsilon_h$ corrections are present. The lagrangian density for D7 probe branes in the Einstein frame is
\be
{\cal L}={\cal L}_{DBI}+{\cal L}_{WZ} = T_7\,e^{\Phi}\sqrt{-\det \hat g_8}+T_7\, \hat C_8\ ,
\ee
with $T_7$ being the tension of the D7-brane and $C_8$ is the RR eight-form potential (see below). Hatted are pull-back quantities in the worldvolume of the probe brane. 
Computing the determinant of the metric (\ref{indmetr}), the lagrangian density of the D7-brane reads
\be
{\cal L}_{DBI} = \frac{e^{\Phi}}{8} \sin\theta
\cos^3{\chi\over 2} \sqrt{-G_{tt} G_{xx}^3 G_{rr}\gke^3}
\sqrt{1+ \left( \frac{G_{\tau\tau}}{\gke} - 1 \right) \cos^2{\chi\over 2}}
\sqrt{1+ \frac{\gke}{G_{rr}}\frac{\chi'^2}{4}}
\label{effL} \ .
\ee

Let us now analyze the WZ term. 
First of all, we take the RR nine-form field strength $F_9$ in the Einstein frame as
$
F_9 = -e^{2\Phi} \star F_1 = -Q_f e^{2\Phi} \star (d\tau +A_{K\!E})
$, 
where we substituted in $F_1$ the form sourced by the background D7-branes. The corresponding $F_9$ is
\be
F_9 = \frac{Q_f}{32}e^{2\Phi}\sin \theta \, \sqrt{\frac{G_{tt}G_{rr} G_{xx}^3 \gke^4}{G_{\tau\tau}}} \cos^2{{\chi\over 2}}\sin\chi\, d^4x \wedge dr \wedge d\chi \wedge d\phi \wedge d\theta \wedge d\xi \ .
\ee
We can define the potential $\hat C_8$ as $F_9=d\hat C_8$ and the corresponding WZ lagrangian density is
\be
{\cal L}_{WZ}\,=\frac{Q_f}{32}e^{2\Phi} R^3 \sqrt{\frac{G_{tt}G_{rr} G_{xx}^3 \gke^4}{G_{\tau\tau}}} \,\left( \cos^4{{\chi\over 2}}\right)\sin\theta
\,\,.
\ee
Rewriting the dilaton by expressing explicitly its value at the horizon, $e^{\Phi(r)} = e^{\Phi_h+\phi(r)}$, and using the definition $\epsilon_h = Q_f e^{\Phi_h}$, we arrive at the total 8 dimensional lagrangian density
\beqa
{\cal L}&=& \frac{T_7 e^{\Phi_h}}{8} e^\phi \sin\theta \, \sqrt{G_{tt}G_{rr} G_{xx}^3 \gke^3} \\
&&\times \left(\cos\frac{\chi}{2} \right)^3 \left[
\sqrt{1 + \frac{\gke}{G_{rr}}\frac{ \chi'^2}{4}}
\sqrt{1+\left(\frac{G_{\tau\tau}}{\gke}-1\right) \left(\cos\frac{\chi}{2} \right)^2}
+ {\epsilon_h\over 4}\, e^{\Phi}\sqrt{\frac{\gke}{G_{\tau\tau}}}\,\cos{{\chi\over 2}}\right] \ . \nonumber
\eeqa
It will be useful to make the following change of variable $\psi(r) = \sin(\chi(r)/2) $ in order to make contact with the notation used in the literature\footnote{This is exactly the same function that is called $\chi = \cos\theta$ that appears in \cite{myers}. Clearly massless embeddings have $\psi =
0$.}. The dimensionless action density, $I_{D7}\equiv S/ Vol({\bf R_3})$, acquires the following form after integrating in $t\in (0,\beta=1/T)$, $x^i$ and $\theta,\phi$ and $\xi$
\beqa
I_{D7}&=&{\cal N}(\epsilon_h) \displaystyle \int dr \, \frac{e^\phi}{r_h^4}\sqrt{G_{tt}G_{rr} G_{xx}^3 \gke^3} \, \tilde\psi^3 \left[
\sqrt{1 + \frac{\gke}{G_{rr}} \left(\frac{\psi'}{\tilde \psi}\right)^2}
\sqrt{1+\left(\frac{G_{\tau\tau}}{\gke}-1\right)\tilde\psi^2}
\right.
\nonumber\\
&& \qquad \qquad \left. +
{\epsilon_h\over 4}e^{\phi}\sqrt{\frac{\gke}{G_{\tau\tau}}} \tilde\psi\right] \ ,
\label{total-L-general}
\eeqa 
where we have defined the short hand notation $\tilde\psi = \sqrt{1-\psi^2}$. Notice than in (\ref{total-L-general}) the integral is dimensionless thanks to the factor $r_h^{-4}$ which has been added by hand for convenience. The prefactor, which gives the right dimension 3 to the action density $I_{D7}$, reads
\beqa
{\cal N}(\epsilon_h) &=& \frac{ 2\pi^2T_7 e^{\Phi_h} r_h^4}{T} N_f'\nonumber\\
 &\equiv&{\cal N}^{(0)}
 \left( 1 + \frac{\epsilon_h}{4} +\frac{3}{32}\epsilon_h^2+ \cdots \right) \ , \label{dimensionsfactor}
\eeqa
with
\be
{\cal N}^{(0)} = \frac{\lambda^{(0)}_h T^3 N_f' }{16} \ .
\ee
Expressing \eqref{dimensionsfactor} in terms of field theoretical quantities was made by means of the relations
$T_{7}^{-1} = (2\pi)^7 g_s l_s^8$, $R^4 = 4\pi g_s N_c l_s^4=g_{YM}^2 N_c l_s^4$, and $\lambda_h = 4\pi g_s e^{\Phi_h} N_c$, as well as \eqref{rhshift}-\eqref{thooftsh}
which will be used extensively during this note.

The equations of motion 
for the embedding can be derived from \eqref{total-L-general}. After inserting the perturbative solutions given above for the metric components and dilaton, the equations of motion can be expanded in powers of $\epsilon_h$. A perturbative solution for the embedding involves an ansatz
 \be
 \psi(r) = \sum_{p=0}^\infty \epsilon_h^p \, \psi_{p}(r) \ ,
 \ee
where each function $\psi_{p}(r)$ solves the equations of motion at a given order $\epsilon_h^p$. As mentioned above, we shall only compute up to $\psi_1(r)$
and consider this as the first order correction to the physics of the probe induced by the flavored background.

\section{Flavor corrections to mass and condensate}\label{sec:masscond}

In the previous section we have displayed an effective gravitational solution where we have decoupled the UV Landau pole by sending it to
$r_*=\infty$. In \cite{D3D7QGP} this solution was shown to provide sensible results for IR quantities of the dual field theory. In particular the
thermodynamics as well as the hydrodynamics \cite{BCTHydro} are found. It seems natural to expect difficulties when addressing questions which
involve a proper treatment of the UV divergences through holographic renormalization. It is the purpose of this section to elaborate on such aspects
for the probe branes.

Central to the discussion is the choice of integration constants. The perturbative expansion of the equation of motion for the embedding profile,
$\psi(r)$, leads to a tower of second-order ordinary differential equations for the components $\psi_p(r), \, p= 0,1,\cdots$. A series solution
around the UV, $r\sim \infty$, depends on two constants, related to the bare mass of the quark $M_q$, and the condensate $\langle \bar \psi \psi
\rangle$ in a way that we will clarify.
The value of the constituent quark mass is a natural boundary condition to choose and, naively, the value of the condensate would be the second condition to impose. However, a generic value of $\langle \bar \psi \psi \rangle$ leads to a divergent solution for the bulk field $\psi_p(r)$ in the IR of the theory. This can be avoided by imposing regularity in the IR as the second boundary condition.

Numerically it is easier to start the integration from the IR towards the UV. Then regularity is satisfied by construction, and the series solution
around the IR limit depends upon a single integration constant, $\xi$, which parametrizes the condition we impose there. For example, consider the
unflavored component $\psi_0(r)$. Depending on whether we are dealing with a {\em black hole} or a {\em Minkowski} embeddings \cite{Mateos:2006nu},
the integration constant can be the insertion angle of the D7 brane when it pierces the horizon $\xi=\psi_0(r_h)$, or the minimum distance
$\xi=r_{min}$ between the probe D7 and the color D3 branes, defined by $\psi_0(r_{min})=1$, respectively.

 Concerning $\psi_1(r)$, the first order correction in $\epsilon_h$ to the embedding profile, it is governed by a second-order, \emph{linear}, non-homogeneous differential equation. As before, we integrate numerically from the IR towards the UV with adequate boundary conditions. For black hole embeddings we chose $\psi_1(r_h)=0$ and for 
 Minkowski embeddings $\psi_1(r_{min})=0$, so that both the anchoring point and the minimal distance are fully encoded in $\psi_0(r)$.
 
Up to the first correction in flavor backreaction the embedding profile has an expansion for large $r$ given by
\beqa
\psi(r) &=& \psi_{0}(r) + \epsilon_h \psi_{1}(r) + \cdots \label{asymprofile}\\
&=& \frac{r_h}{r} 
\left[ m_0 + \left(\frac{r_h}{r}\right)^2c_0 +\epsilon_h\left( m_{1} - \frac{m_0}{6}\log\frac{r}{r_h} + \left( \frac{r_h}{r}\right)^2 \left( c_1 - \frac{ 5 c_0 }{6} \log \frac{r}{r_h} \right)
\right)+ \cdots \right]\ , \nonumber
\label{uvexp}
\eeqa
where the constants $m_0$, $m_1$, $c_0$ and $c_1$ all depend on $\xi$ and can be extracted from a fit to the numerical integration. We see the
appearance of logarithmic terms in the expansion. A similar effect has been observed in various situations before. In
\cite{Karch:2006bv,HoyosBadajoz:2010td} it was seen to arise from a coupling to the Ricci scalar in four dimensional space. Our situation here is more
akin to the one in \cite{Albash:2011nw}. Observe that we obtain logarithms contributing both to the mass term and to the condensate, so the proper
identification of lagrangian mass and condensate are suspect.  However, we shall argue later that, for such an embedding, the quark bare mass is
indeed given by the sum\footnote{In fact, one can perform a reparameterization of the radial coordinate $r \to\tilde r = r(1 + {\cal O}(\epsilon_h)
+\cdots)$, such that the logarithmic term that shifts $m_1$ disappears.} $m= m_0 + \epsilon_h m_1.$

Conversely, we may be interested in seeing how the D7 profile is modified by the addition of flavors for a fixed value of $m$.
This amounts to shifting $\xi \to\xi' = \xi - \epsilon_h \xi_1$, in such a way that $m(\xi') = m_0(\xi)$ stays invariant.
This is important, for example, in order to deduce how the presence of flavors modifies the constituent mass, $M_c$ \footnote{In \cite{herzog1} the
constituent mass is denoted rest mass $M_{rest}$.}.
 This is only defined for Minkowski embeddings, and is obtained from the Nambu-Goto action for a fundamental string that stretches from the horizon up to the point of nearest approximation of the D7 brane to the origin. 
The relevant calculation is 
\beqa
M_c(\epsilon_h) &=& \frac{1}{2\pi \alpha' } \int_{r_h}^{r_{min}} e^{\Phi/2}\sqrt{-G_{tt}G_{rr}} \, dr dt \nonumber\\
&=& \frac{e^{\Phi_h/2}}{2\pi \alpha' }
\left[ 
\left(1-\frac{3\epsilon_h}{8}\right)(r_{min} - r_h)+\frac{\epsilon_h}{2} r_{min} \log\frac{r_{min}}{r_h} + \cdots \right]
 \label{constmass} \ .
\eeqa
 In the flavorless limit $\epsilon_h= 0$, the above expression allows for a physical definition of the bare quark mass. In that case, the UV regime is
safe and conformal and, thus, in the limit of very high masses as compared with the plasma temperature $T$, the constituent mass $M_c$ and the quark
mass $M_q$ should become the same.
 From (\ref{asymprofile}) in the unflavored limit, we see that the definition of the minimum distance leads to $r_{min} = m r_h +{\cal O}(1/r) $. Hence
 \be\label{baremass} 
 M_q = \lim_{r_{min}\to \infty} M_c(0) = \lim_{r_{min}\to \infty} \frac{e^{\Phi_h/2}}{2\pi \alpha' }(r_{min} - r_h) = \frac{1}{2}\sqrt{\lambda_h^{(0)}} \, T\, m \ .
 \ee
This result also holds in the limit $T\to 0$, as can be seen from the exact supersymmetric embedding: $m r_h = r \psi(r)$.
 
When $\epsilon_h$ is not zero this nice matching is lost. Naively we would expect the large mass limit to give $M_c\to \frac{1}{2}\sqrt{\lambda^{(0)}_h} T
(m_0+\epsilon_h m_1)$, since a very heavy quark should decouple both from thermal fluctuations as well as from the massless quarks in the sea. However
taking the limit of large $r_{min}$ in (\ref{constmass}) gives a divergent result. This is, as adverted, one of the drawbacks that arise from
truncating the perturbative expansion in powers of $\epsilon_h$. Hence $M_c(\epsilon_h)$ cannot be trusted beyond $r_{min}\sim r_{\Lambda} = r_h e^{1/\epsilon_h}$. For an
$\epsilon_h\sim 0.24$ (which would correspond to a realistic regime with $\lambda_h \sim 6\pi$ and $N_f=N_c=3$) this gives a safe range for $r_{min}/ r_h\sim m \lesssim {\cal O}(10)$. In this range we must anyway resort to numerical analysis. 
\begin{figure}[th!]
\begin{center}
\includegraphics[scale=0.8]{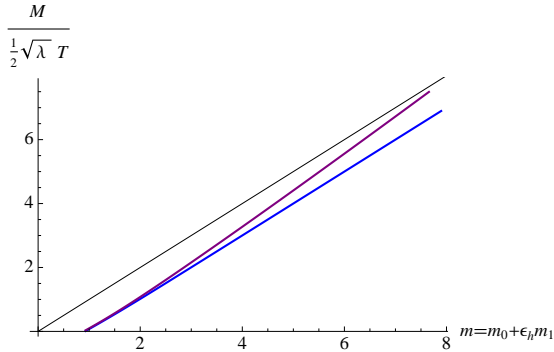}
\caption{\em In this figure we plot $M/(\frac{1}{2} \sqrt{\lambda}T)$ for different $M$'s. 
For the black diagonal line $M=M_q$ the lagrangian quark mass. In blue we see $M=M_{c}^{(0)}$, the constituent mass in the unflavored thermal background.
It vanishes for $r_{min} \sim 0.92$. The purple curve is for $M=M_c$, the constituent mass in the flavored thermal background. It seems to 
cross the black line at a finite value of $m$. This is an artifact of the approximation and does not happen for the restricted range of $m$. In a realistic model with a controllable UV completion both blue and purple curves should approach the black line for $m\to \infty$
 }
\label{constituentmass}
\end{center}
\end{figure}

This means in the context of the present analysis, we should only consider quark masses bound as $M_q = \frac{1}{2}\sqrt{\lambda_0}T m \lesssim {\cal O}(20\, T) \sim 3.5$ GeV
(taking for $T$ the critical temperature $T_c = 175$ MeV as the lowest possible physical value).
In the plot of
figure \ref{constituentmass} we see the influence of the massless unquenched flavors on $M_c$ of the quenched probe in the advocated restricted range. The generic lesson is that fundamental degrees of freedom contribute more than adjoint ones to the constituent mass.

\subsection{Holographic renormalization, quark mass and condensate}\label{renormandcond}

In this section we will justify the aforementioned identification of the bare mass. We will moreover provide an expression for the quark condensate.
First of all, let us deal with the on-shell action. After inserting the asymptotic expansion (\ref{asymprofile}) into the action
(\ref{total-L-general}), we obtain quartic and quadratic divergences that can be tamed with a standard counterterm first proposed in
\cite{Karch:2005ms} and, in a similar context to ours, in \cite{Filev:2011mt}
 \be
I_{ct,1} = {\cal N}(\epsilon_h) \left. \sqrt{\hat\gamma}\, e^{\phi} \frac{R^4}{4 r_h^4} \left(1 + \frac{\epsilon_h}{6}\right) \left(1-\psi(r)^2\right)^2
\right\vert_{r_c} \ ,
\label{counter1}
 \ee
with $\hat\gamma$ the determinant of the induced metric on the $r=r_c$ hypersurface, which is a convenient (large) cutoff. The subtraction $I_{D7}-
I_{ct,1}$ removes all power like divergences, but leaves a logarithmic one. This last can be
eliminated with an additional counterterm of the form
\be
I_{ct,2} = {\cal N}(\epsilon_h) \frac{\epsilon_h}{8 }r^4_h\log \frac{r_c}{r_h} \label{counter2} \ .
\ee
It is remarkable that this divergence is proportional to $r_h$ and, hence, only pops up upon turning on the temperature. In the supersymmetric case,
the addition of (\ref{counter1}) is just enough. It is unclear to us what can be the source of this new divergence, although there might be several
possibilities (see for example \cite{vanRees:2011fr,Papadimitriou:2011qb}). 
The important fact to stress is that, unlike the case in \cite{Albash:2011nw}, this counterterm is totally independent of the details of the probe
brane profile. Therefore, it will not affect the computation of the condensate.

Let us now proceed to compute the 1-point function. One would naively consider $\langle \bar \psi \psi \rangle =T \partial I_{ren}/\partial m$ with $m= 
m_0+\epsilon m_1$ as a correct prescription for the condensate. 
In this sense the renormalized action $I_{ren} = I_{bulk}- I_{ct,1} $ is not useful, as it does not exhibit the form of a product of {\em source} times {\em v.e.v.} $\sim \int J \langle O\rangle$ where $J\sim m_0 + \epsilon_h m_1$.
As very well explained in \cite{myers}, the point is that both
$c_0$ and $c_1$ depend implicitly on $m_0$ and $m_1$ through the fact that all of them are controlled by a single parameter: either $\xi=r_{min}$ for
Minkowski, or $\xi=\psi_h$ for black-hole embeddings. Hence it will be convenient to think of $I_{ren}$ as parameterized by $\xi$. We compute
$I'_{ren}(\xi)$, yielding 
\be
I'_{ren}(\xi) = I'_{D7}(\xi)- I'_{ct,1}(\xi) = \left.\frac{\partial{ \cal L}}{\partial\psi'} \frac{\partial\psi}{\partial\xi} \right\vert^{r_c}_{r_{min}}- \left.\frac{\partial I_{ct,1}}{\partial\psi} \frac{\partial\psi}{\partial\xi} \right\vert^{r_c} \ ,
\ee
where $r_c$ is a cutoff radius, use of the Euler Lagrange equations of motion has been made, and an integration over all coordinates but $r$ has been understood in
$\cal L$. As said before, $I_{ct,2}$ does not contribute to this calculation since it does have any dependence on the embedding profile.
The contribution at $r_{min}$ vanishes, and after adding up both terms the final result organizes in a neat form
\be
I'_{ren} = - {\cal N} \left[ 2c_0 + \epsilon_h \left(2 c_1 + \frac{7}{6}c_0\right)\right] (m_0' + \epsilon_h m_1')+\cdots \ .
\ee
Using the chain rule this gives the holographic dictionary we are after, if we interpret the bare quark mass as $M_q = \frac{1}{2} \sqrt{\lambda_h^{(0)}}T m$ with $m=(m_0 + \epsilon_h m_1)$
\be
\langle {\bar\psi \psi}\rangle =\frac{\partial I_{ren}}{\partial M_q}=- \frac{\sqrt{\lambda_h^{(0)}}T^3 N_f'}{8} \left[ 2c_0 + \epsilon_h \left(2 c_1 +\frac{7}{6} c_0\right)\right]\, .
\label{condensate}
\ee

\begin{figure}[htbp]
\begin{center}
\includegraphics[scale=0.55]{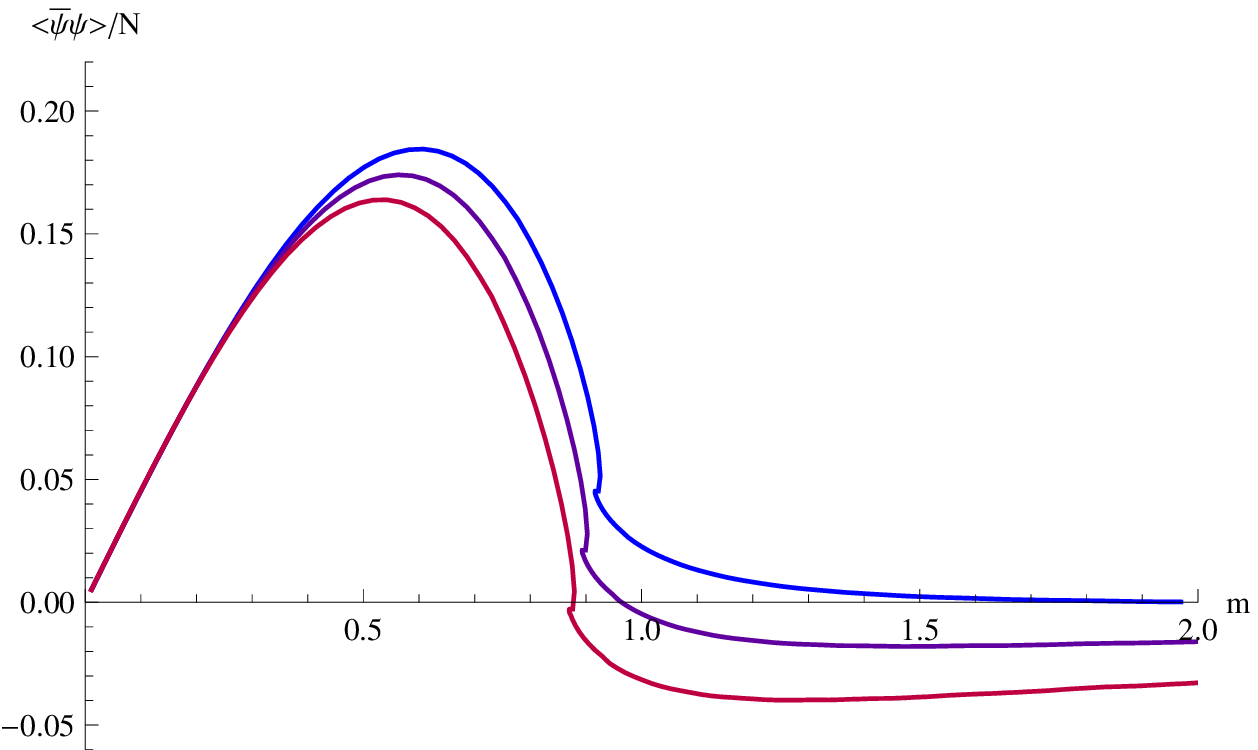}
\hspace{0.5cm}
\includegraphics[scale=0.55]{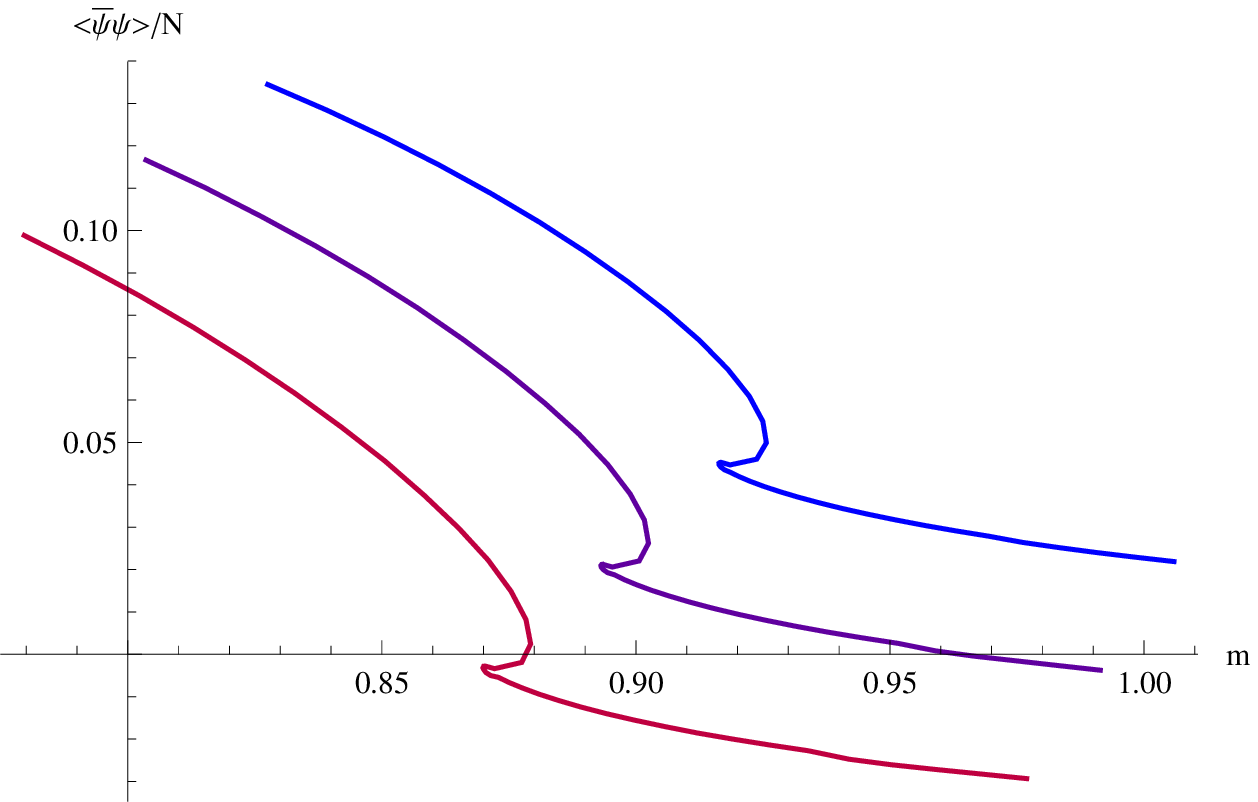}
\caption{\em Here we observe that the effect of the flavor mass density on the heavy quark condensate is to lower its value. From top to bottom the curves correspond to values of $\epsilon_h = 0, 0.2$ and 0.4 respectively. On the right hand side we take a closer look at the transition region and observe that the critical mass $m_{cr}$ is reduced, as can be deduced from the Maxwell equal-area construction}
\label{condensflavor}
\end{center}
\end{figure}

In figure \ref{condensflavor} we have plotted the values of the condensate versus the mass. It is intriguing to observe that the effect of the massless backreacted flavors is to lower the curve in such a way
that the condensate changes sign at a given value of the mass, whose position depends on $\epsilon_h$. A similar effect occurs upon turning on a magnetic field $B$ as found in \cite{Albash.et.al.,JohannaReneJon}. Unlike those papers, here we cannot see if a spontaneous chiral symmetry breaking occurs for high values of $\epsilon_h$, as this parameter is perturbative in our approach. 
The presence of unquenched matter also shifts the point of the meson-melting transition, such that for fixed $M_q$ the transition happens at a higher temperature\footnote{Notice that, as compared with 
\cite{myers}, our $\lambda=4\pi g_s$ is a factor of $2$ larger. Expression \eqref{baremass} however is the same because in \eqref{uvexp} we have absorbed a factor of $1/\sqrt{2}$ in $m_0, m_1$, as well
as a factor $1/(2\sqrt{2})$ in $c_0,c_1$. This explains the discrepancy in the meson melting point, which is at $m_0=0.923$ for us instead of being at $1.305$ for them.}.

\section{Conductivity and transport}\label{sec:conduc}

The presence of massless fundamental degrees of freedom in the background has an impact on the transport properties of the probe brane. Fluctuations
of the fields defined in the worldvolume of the probe D7 are sensitive to the $\epsilon_h$ corrections in the background fields (\emph{i.e.}, metric
and dilaton), and so is the response of the system to these perturbations. The relations between these responses and the fluctuations sourcing them
define the transport coefficients. A paradigmatic example is the (microscopic) conductivity, which relates, with the aid of linear response theory,
the microcurrent induced in a system to the fluctuation of the electric field that causes it. In a homogeneous and isotropic medium this relation reads
$
\delta \vec J(\omega) = \sigma_{micro}(\omega) \delta \vec E(\omega) 
$, 
where $\omega$ is the frequency of the fluctuation.

There are examples in the literature where the calculation of $\sigma_{micro}$ has been performed in different backgrounds. We refer to (some of)
those works for details of the calculation that we only sketch in the this note. We will restrict our study to the DC limit of the conductivity an
alternative approach can be used, as devised in \cite{arXiv:0705.3870}, with a clear physical significance. This approach implies a macroscopic
calculation of the conductivity in which one mimics an Ohm's law experiment. Instead of studying fluctuations of the electric field in a setup where
no background $\vec E$ is present, one considers a worldvolume gauge field of the form
$
A= A_0(r) dt + \left( E_x t+A_x(r) \right) dx \ .
$
In this expression $E_x$ is a constant electric field, that we align with the $x$ direction. The radial dependent fields $A_0$ and $A_x$ determine holographically the baryon density, $n_q$, and charge current, $\langle J_x \rangle$, in the field theory \cite{arXiv:0705.3870}. From Ohm's law we define the macroscopic conductivity as
$
\langle J_x \rangle = \sigma_{macro} E_x \ .
$
Using the membrane paradigm it has been shown that the macroscopic and microscopic conductivities are related as
\be
\lim_{E_x\to0}\sigma_{macro}(E_x) = \lim_{\omega\to0} \sigma_{micro} (\omega) \equiv \sigma_{DC} \ ,
\ee
but the macroscopic calculation is much neater and direct. 

\begin{figure}[htbp]
\begin{center}
\includegraphics[scale=0.8]{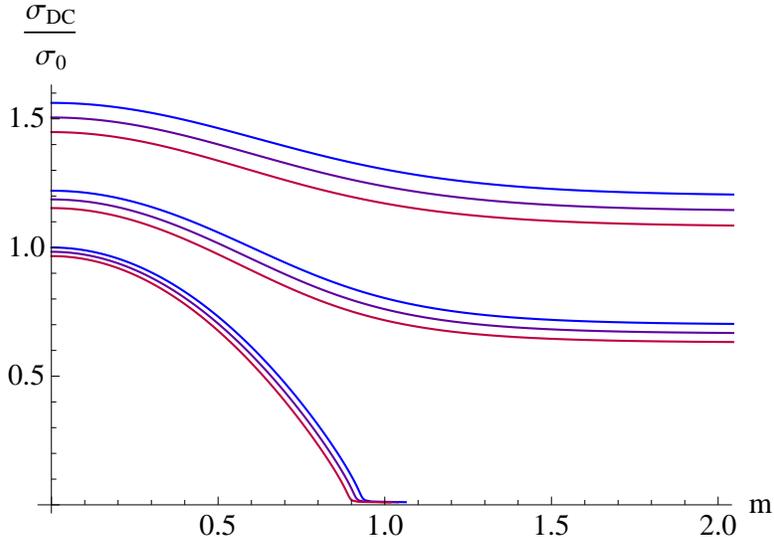}
\caption{\em $\sigma_{DC}/\sigma_{DC}^{(0)}$ as a function of the (renormalized) constituent quark mass parameter $M_q$ for values of
$\epsilon=0,0.2,0.4$ (from blue to red) and $\tilde d=0.01,0.7$ and $1.2$ from bottom to top.}
\label{conductsfig}
\end{center}
\end{figure}

The calculation can be done for black hole embeddings only, since Minkowski embeddings represent an insulating phase where the conductivity vanishes
due to the discrete spectrum. Following the standard procedure described in \cite{arXiv:0705.3870}, we obtained the following result:
\be\label{conductivityresult}
\sigma_{DC} = \sigma_{DC}^{(0)}\, \sqrt{ \left(1 - \displaystyle\frac{\epsilon_h}{6}\right)\, (1-\psi_h^2 )^3 + 
(\tilde d^{(0)})^2\left( 1 - \frac{\epsilon_h}{4}\right) \ ,
}
\ee
where $ \psi_h = \psi (r_h)$ and we have defined
\be
\sigma_{DC}^{(0)} = \frac{N'_f N^{(0)}_c}{4\pi} T \ , \qquad \tilde d^{(0)} = \frac{8 n_q}{N_f' N^{(0)}_c\sqrt{\lambda_h^{(0)}} T} \ .
\ee
To plot expression \eqref{conductivityresult} we do not need to calculate the embedding profile in the presence of finite baryon density, if we want
to plot the conductivity as a function of $\psi_h$. This step becomes mandatory, though, if we want to trade the IR quantity, $\psi_h$, for the
physical UV one, $m$, as the independent variable in fig. \ref{conductsfig}.\footnote{
Obtaining the modified embedding profile goes like in section \ref{sec:probes}, but with a non trivial gauge field $A=A_0(r) \dd t$. The presence of this field in the D7 action shows up in two places. The first one is in the DBI part of the action, which expands from $\sqrt{-\det g}$ into $\sqrt{-\det (g+2\pi \alpha'F)}$. As a consequence, the last square root in equation (\ref{effL}) should be modified as follows
\be
\sqrt{1+ \frac{\gke}{G_{rr}}\frac{\chi'^2}{4}} \to \sqrt{1+ \frac{\gke}{G_{rr}}\frac{\chi'^2}{4}-\frac{(2\pi \alpha')^2 A_0'^2 }{|G_{tt}|G_{rr}}} \ .
\ee
The second contribution has to be considered when the unquenched fundamental matter has a net charge \cite{D3D7QGPb}. Then, in the Wess-Zumino part of the probe action there is a term coming from $\hat C_6 \wedge (A_0' \dd t \wedge \dd r)$, where $\hat C_6$ is the pullback of the background $C_6$ RR potential to the worldvolume of the probe brane. This gives a term proportional to $J A_0'$, with $J$ the function taking into account the effects of the backreaction of charged D7 branes (see reference \cite{D3D7QGPb} for more details on this function). We will stick to the case in which the background is not charged, therefore $J=0$ and the WZ contribution does not get modifications.
}

 As pointed out in \cite{arXiv:0705.3870}, two different mechanisms for the conductivity can be identified in (\ref{conductivityresult}). One of them
comes from the presence of a density of free charge carriers proportional to the density charge. Its contribution gives rise to the $\tilde d^{(0)}$
term under the square root. The other one remains even in the charge-less limit $\tilde d^{(0)}\to 0$ and is proportional to the
temperature-dependent factor $\sigma_{DC}^{(0)}$, signaling a thermal origin. The interesting fact pointed out in \cite{arXiv:0705.3870} is the
presence of the modulating factor $(1-\psi_h^2)^{3/2}$ which suppresses this contribution for high masses ($\psi_h \to 1$). This points to a
Schwinger-like pair production of charged carriers as the mechanism behind this effect. 

In \eqref{conductivityresult} the correction to these terms due to the backreaction of flavor degrees of freedom is a decreasing one. For the term
depending explicitly in $\tilde d^{(0)}$, the electric field accelerates the valence charge carriers, and the drift of charged matter gives rise to a
macroscopic current. As they move, they scatter against the bulk adjoint and fundamental degrees of freedom that can absorb momentum and energy from
these carriers and, therefore, slow them down. The negative correction $(1-\epsilon_h/4)$ means that the fundamental degrees of freedom have a larger
cross section than the gluonic ones they have replaced. As we will see, this is in agreement with the enhancement of the drag force on a heavy test
quark.

Concerning the first term, we see that the correction is also negative $(1-\epsilon_h/6)$, whereas we would expect the pair creation to be enhanced by the presence of charged massless fundamental matter that can run inside the loops.
Here we have to stress the importance of the comparison scheme selected. We have chosen to keep invariant the energy density, and therefore the number of colors 
is diminished according to \eqref{Ncshift}. Had we decided instead to compare two theories with the same number of colors, and just additional fundamentals, the correction to the first term would have been $(1+\epsilon_h/3)$, hence positive in agreement with the aforementioned expectations. From the fact that the net correction is negative, we conclude that at fixed number of degrees of freedom the fundamental matter contribute less to the pair-production than the adjoint matter.

In figure \ref{conductsfig} we plot the conductivity as a function of the quark mass $m=(m_0+\epsilon_h m_1)$ for various values of the density $\tilde d^{(0)}$. 
For $\tilde d^{(0)}\to 0$ the curves drop to zero conduction (insulator) at a point that drifts from $m\sim 0.92$ downwards with $\epsilon_h$. This is
consistent with the shift in the meson melting transition point observed in fig. \ref{condensflavor}. We observe that at large mass the pair creation
term is negligible and the conductivity approaches a constant $\tilde d(1-\epsilon/8)$.

\section{Quark potential}\label{sec:qpotential}

It is well known that one can extract the potential energy of an external quark-antiquark pair from the expectation value of a Wilson loop operator
\be
\langle W({\cal C}) \rangle = A(L) e^{-T V(L)} \ ,
\ee
for a rectangular loop which is an infinite strip $T\to \infty$. Here $L$ is the quark-antiquark separation in Minkowski space. The classical
calculation in \cite{Maldacena:1998im,Rey:1998ik,Rey:1998bq,Brandhuber:1998bs} can be easily adapted to the present setup. Since the only
dimensionful parameter is the temperature, we expect on general grounds a screened Coulomb potential of the following form
\be
V(L) = -\frac{Q(\epsilon)}{L}\left( 1 + c(LT)^4 + \epsilon f(LT)+\cdots \right) \ ,
\ee
where $c$ is a constant \cite{Brandhuber:1998bs}. 
The idea here is to parametrize both $L$ and $V$ through the hanging string maximum depth $r_0$
\beqa
L &=& \int_{r_0}^{r_{\text{min}}(\epsilon_h,m)}{dr \;\frac{R^2 \sqrt{r_0^4-r_h^4}}{\sqrt{(r^4-r_0^4)(r^4-r_h^4)}}}\left(1+ \frac{\epsilon_h}{8}\left(1-4\frac{r^4-r_h^4}{r^4-r_0^4}\,\log \frac{r}{r_0} \right) + \cdots \right) \ ,
\\
E &=&
 \int_{r_0}^{r_{\text{min}}(\epsilon_h,m)} dr\left[ \frac{\sqrt{r^4-r_h^4}}{\sqrt{r^4-r_0^4}}
 \left(
 1+ \frac{\epsilon_h}{8}
 \left(1+4\, \log \left[\frac{r}{r_h}\right]-4\frac{r_0^4-r_h^4}{r^4-r_0^4}\, \log \left[\frac{r}{r_0}\right]\right)
 \right)\right.
\\ &&
\qquad \qquad \qquad \qquad \left. -\left( 1+ \frac{\epsilon_h}{8}\left(
 1 +4\log \left[\frac{r}{r_h}\right]\right)\right) +\cdots\right] \ .
\nonumber
 \eeqa
In this expression, $r_{min}(\epsilon_h,m)$ is given by the point of minimum radius for the probe D7 for a given value of the bare mass parameter $m$. It has to be obtained numerically by shooting
from different minimal values of $r_{min}$ until, in the UV, the desired value of $m= m_0 + \epsilon_h m_1$ is hit.
\begin{figure}[htbp]
\begin{center}
\includegraphics[scale=0.5]{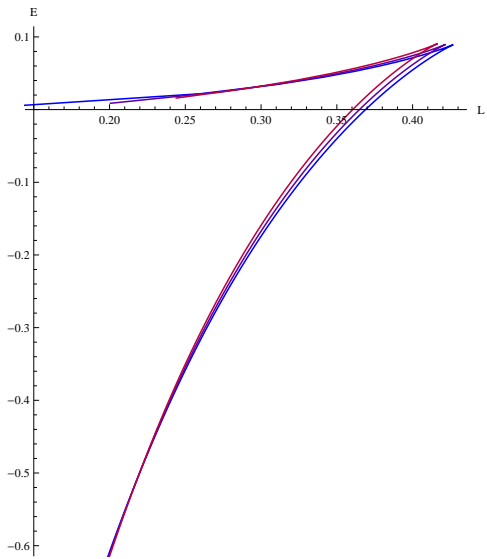}
\caption{\label{qqpotfig}\em{ The quark-antiquark potential $E_{q\bar q}$ as a function of the interquark separation $L$. The usual plot shows a
turning cusp point where is ceases to exist. Before that, the configuration with two disconnected strings dominates whenever the energy changes sign.
This defines a screening length $L_{sc} = 0.37$, beyond which the interquark potential flattens. As we see in the figure, the main effect of the
flavors is to decrease this length. The plots are for $\epsilon_h = 0, 0.25$ and $0.5$ from blue to red.}}
\end{center}
\end{figure}

In figure \ref{qqpotfig} we show the curves that correspond to the $q\bar q$-potential for three values of $\epsilon_h$. 
The curves have typical shape obtained in \cite{Rey:1998bq,Brandhuber:1998bs}, which have been interpreted as the thermal screening of the plasma over
the Coulomb potential. They exhibit a cusp at a certain separation $L_{max}$, beyond which the $\cup$-shaped string embedding ceases to exist. This
length is however larger
than the dissociation or screening length $L_{sc}$, were the curve $E(L)$ changes sign. Beyond this point, the dominant saddle point is that of a pair of disconnected strings
hanging vertically down to the horizon.

In the figure, we observe how the addition of massless flavors moves the curve towards the left. Hence the onset of a flat potential occurs at a
closer separation that in the unflavored case. This behavior is in agreement with the result obtained in \cite{Bigazzi:2008zt} for the same
calculation in the Klebanov-Witten model at zero temperature. We also notice the truncation of the curves in the left plot of their fig. 3, due to the
horizon which triggers the screening by string splitting (the saddle point with two hanging strings becomes favorable). Hence the usual double
valuedness leading to the cusp is in the unphysical region. 
This is a direct consequence of the fact that, keeping the temperature fixed, demands rising the value of the horizon radius, $r_h$, in \eqref{rhshift}.

\section{Drag force}\label{sec:dragforce}

Let us now turn our attention to the issue of energy loss of partons in this medium.
In the real world, the QCD quark-gluon plasma appears to be strongly coupled \cite{shur}.
Moreover a very efficient mechanism of energy loss must be at work in order to explain the phenomenon of jet quenching \cite{Baier}. 
There are two main ways of addressing the computation of this coefficient within the holographic paradigm.
One makes use of an eikonal approximation at high energy yielding a non-perturbative definition of $\hat q$ as the coefficient of $L^2$ in an almost light-like Wilson loop with dimensions $L^{-}\gg L$. The implementation of this proposal in the string theoretical framework was done for the first time in \cite{liu}.

\begin{figure}[ht]
\centerline{
\begin{tabular}{c@{\hspace{1.5cm}}c}
 \includegraphics[scale=0.55]{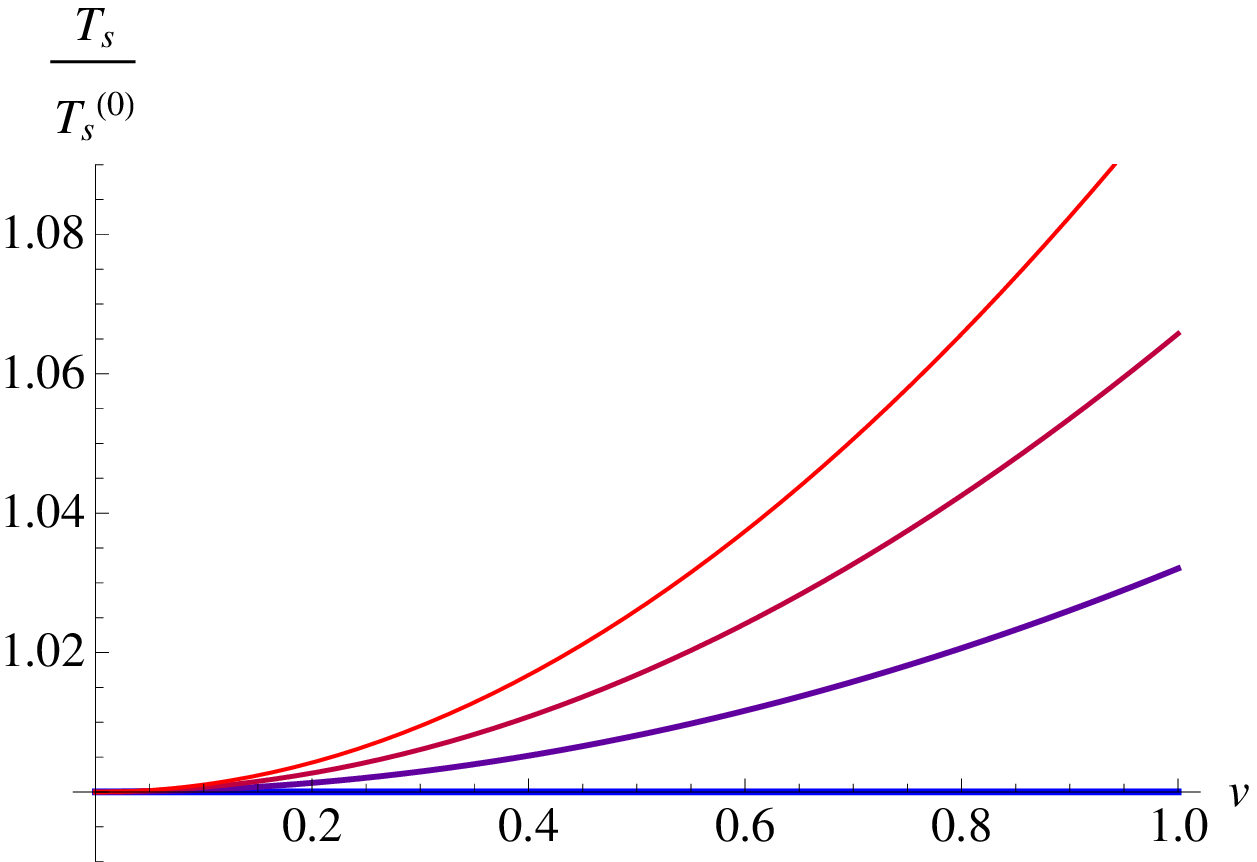} & \includegraphics[scale=0.55]{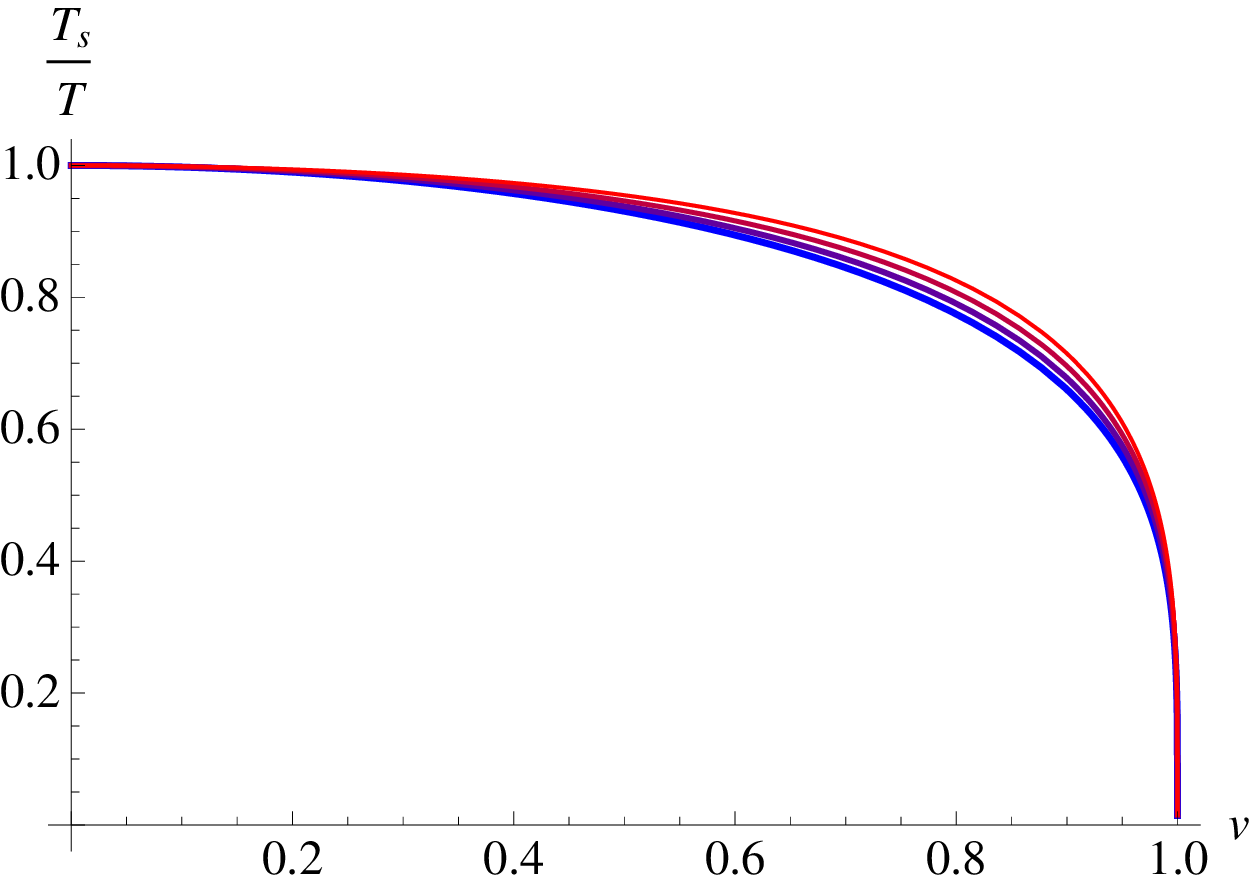}
\end{tabular}
}
\caption{\em{ \label{fig:temps}On the left we plot the worldsheet temperature as the quark moves inside the flavored plasma, relative to the unflavored case. On the right we see the modification of this worldsheet temperature over the bulk temperature is suppressed in the region $v\to 1$ by the $\gamma$ factor.
The different curves correspond 
to increasing values of $\e_h=0,0.25,0.5,0.75$, from blue towards red.}}
\end{figure}

The other involves the picture of a quark traveling through the medium modeled entirely within a string theoretic framework.
 A parton moving inside the plasma is described by a macroscopic string attached to a probe flavor brane placed in the background that contains a black hole.
 The interaction with the medium is encoded in the string profile that dangles down the bulk and extends infinitely along the horizon. One searches for solutions for which the dynamics of the string endpoint, hence the quark, can be compared with a solution to the drag force equation
 \be
 \dot p = -\mu p + f \label{dragforce} \ ,
 \ee
 where $\mu$ is the friction coefficient, related to the mean decay time $\mu = 1/\tau$, and $f$ is an external force on the quark.
 As very neatly explained in 
\cite{herzog1}, there are two simple solutions that one can readily device. The first one is stationary $\dot p=0$ as a consequence of the constant
worldsheet electric field that
pulls the string endpoint
\beqa
X^1=vt+x(r),\quad X^i=0, \quad i=2,3\ .
\eeqa
The string description of the dragging phenomenon embodies interesting predictions. Among them, an intriguing one is the appearance of a
black hole in the pullback metric, for finite velocities of the quark. The horizon sits at a different radial position than that of the bulk metric,
$r_s=r_h \sqrt \g $. As a consequence of this, the worldsheet and bulk Hawking temperatures differ by a velocity dependent factor $T_s(v) =
T/\sqrt{\gamma}+ \cO(\e_h)$. This worldsheet black hole is at the origin of the thermal fluctuations that are responsible, in the present setup, for
the transverse momentum broadening of the heavy probes. Moreover, as remarked in \cite{iancu,GKMNdrag}, this is the temperature that should be used in
the relativistic version of the Einstein relations\footnote{As we will explain in the next section, we checked that $T_s$ is indeed the temperature
appearing in the Einstein relation for the correlators, even in the backreacted background, by using the Schwinger-Keldish formalism.}. 
A general expression for arbitrary backgrounds can be found in equation \eqref{genTs}. The flavor modification factor for this quantity is as follows:
\be
T_s = \frac{T}{\sqrt\gamma}\left( 1 + \frac{\epsilon_h}{8}v^2 + \frac{\epsilon_h^2}{384} v^2(8-3v^2)+\cdots\right)\, .
\label{wstemp}
\ee
The worldsheet temperature in the flavored background is higher than
the supersymmetric limit (but still lower than the bulk temperature, as it is expected and shown in figure \ref{fig:temps}).

%

To keep the velocity constant, the momentum absorbed by the charged quark, sitting at the string end-point, is dissipated into the bulk and through the horizon by the
trailing part of the string, at a constant rate.
It is easy to work out a general expression for the drag force in an arbitrary (dilatonic) background (see also \cite{Herzog:2006se} and \cite{GKMNdrag}).
\be
f= \left. -\pi^1_x \right\vert_{r=r_s} = \left.\frac{e^{\Phi/2}}{2\pi\alpha'} \sqrt{G_{xx}|G_{tt}|} \right\vert_{r=r_s} \ .
\ee

\begin{figure}[ht]
\centerline{
\begin{tabular}{c@{\hspace{1.2cm}}c}
 \includegraphics[scale=0.6]{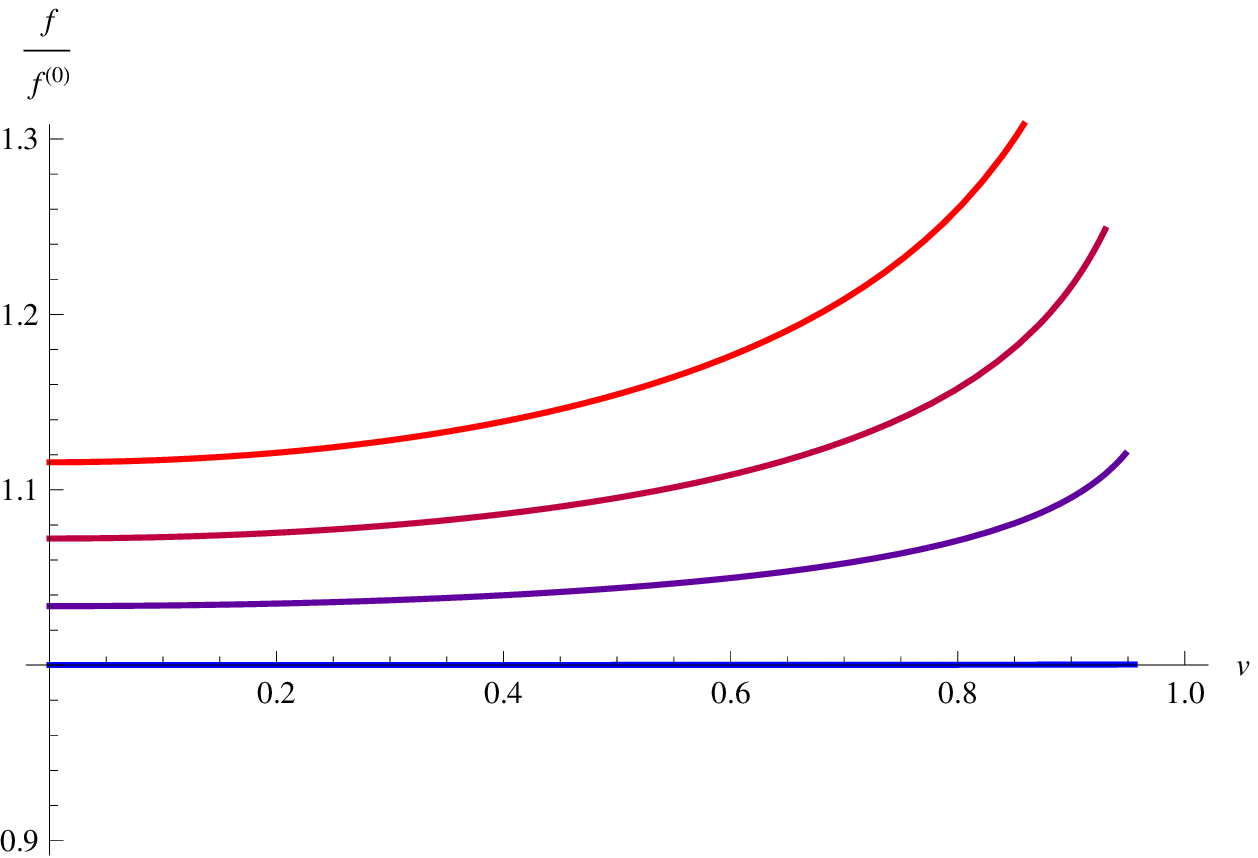} & \includegraphics[scale=0.6]{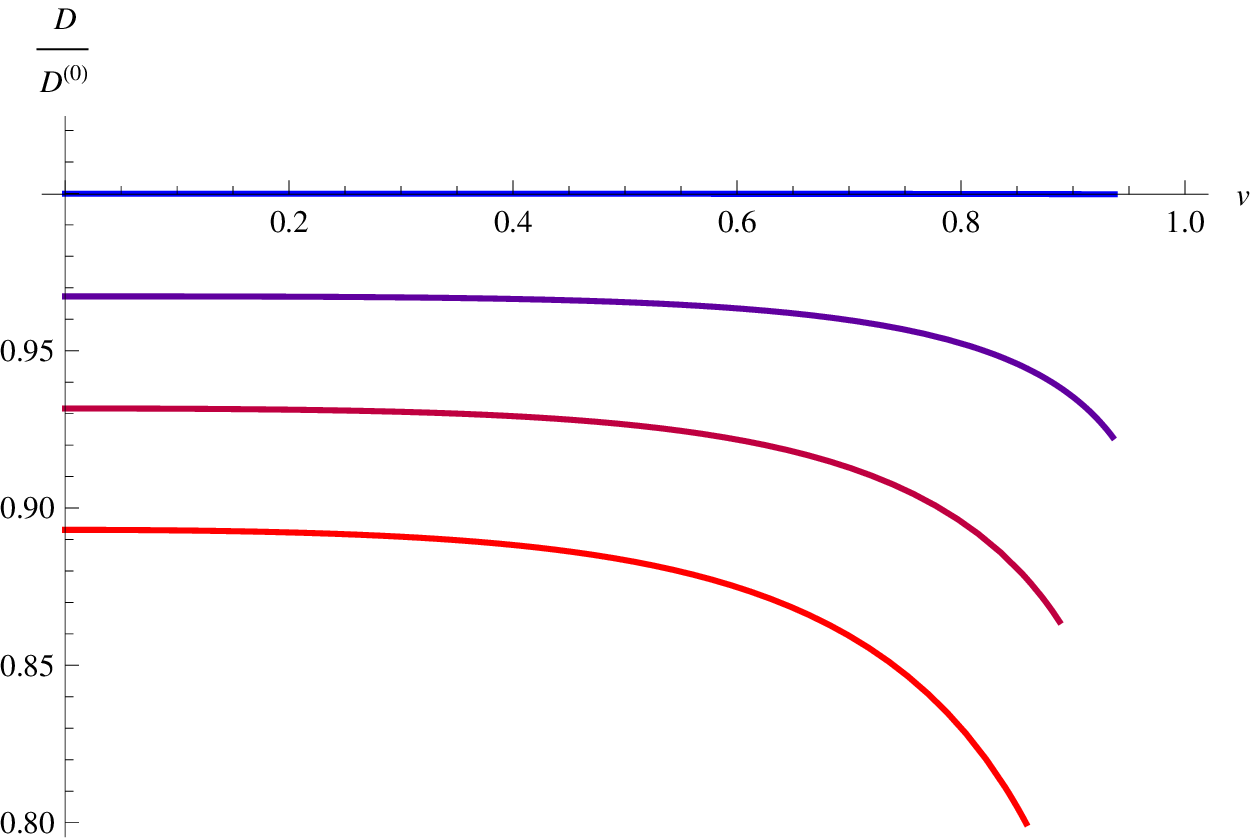}
\end{tabular}
}
\caption{\label{fig:drag} \em{On the left (right), the ratio of the drag force (diffusion constant) with respect to the value it takes in the unflavored background is plotted as a function of the velocity, for different values of the perturbative
parameter $\e_h=0,0.25,0.5,0.75$.}} 
\end{figure}

As will be the case for quantities that do not need the D7 brane profile, we can give a result up to second order in $\epsilon_h$
\beqa
f &=& \frac{\pi\sqrt{\lambda^{(0)}} T^2}{2}\gamma v \left( 1+ \frac{\epsilon_h}{8}(1-\log(1-v^2))+
\frac{\epsilon_h^2}{384}\left(15-14\log(1-v^2) \right.\right.
\nonumber\\
&&\hspace{2cm} ~\left.\left. + 9 \log^2 (1-v^2) + 12\log_2 v^2 \right) \rule{0mm}{5.5mm}\right)+\cdots
\label{dforce} \ ,
\eeqa
 with $\gamma=1/\sqrt{1-v^2}$.
Postulating a relativistic dispersion relation of the form $p = M_{kin}\gamma v$ for some kinetic mass $M_{kin}$, and inserting in 
\eqref{dragforce} (with $\dot p=0$ for a stationary solution),
the above calculation of the drag force implies knowledge of the joint product of $\mu$ and $M_{kin}$
\be
\mu M_{kin} = \frac{f}{\gamma v}\, . \label{muMkin}
\ee
In the unflavored case, dual to $\cN=4$ supersymmetric Yang Mills theory, a value of $\mu^{(0)}$ was obtained that was independent both of the mass and of the velocity of the quark
\be
 \mu^{(0)} M^{(0)}_{kin} = \frac{\pi}{2}\sqrt{\lambda^{(0)}}\, T^2 \ ,
 \label{apdlkkld}
\ee
a fact that contradicts the weak coupling prediction \cite{Moore:2004tg}. From \eqref{dforce} and \eqref{apdlkkld} we see that, in the flavored
background, $\mu$ acquires a logarithmic dependence on the velocity. The product $\mu M_{kin}$ is related to the quark diffusion constant:
\be
D= \frac{T_s}{\mu M_{kin}}\,,
\ee
where, as mentioned above, $T_s$ is the correct temperature to be used here. Of course this is only true if the Langevin equation is a realistic
description of the difusion phenomenon. In fig. \ref{fig:drag} we see the modification of this constant with the flavor density $\epsilon_h$. 
In summary, fundamental flavors contribute more than adjoint matter to the drag force and they inhibit more the diffusion
\beqa
D &=& \frac{2}{\pi T\sqrt{ \lambda^{(0)}\gamma}}
 \left[ 1- \frac{\epsilon_h}{8} \left(\frac{1}{\gamma^2}-\log(1-v^2)\right) - \frac{\epsilon_h^2}{384 \gamma^2} \right.
 \\
& & \left. \times
\left( \rule{0mm}{5mm}
3 - 4\gamma^2 + 10\gamma^4 + (6\gamma^2 - 8\gamma^4)\log(1-v^2) + 3\gamma^4\log^2(1-v^2) + 12\gamma^4 \log_2 v^2
\right)+\cdots \right] .
\nonumber
\eeqa

The friction coefficient $\mu$ is related to the relaxation time $\tau =1/\mu$. In order to disentangle $\mu $ from $M_{kin}$ the authors of
\cite{herzog1} propose to consider a different kind of experiment, in which the string momentum decays exponentially to zero in the absence of drag. 
Then $f=0$ in (\ref{dragforce}) implies that $p(t) = p(0)e^{-\mu t} $ for the probe quark. In the string picture this involves a quasinormal mode
computation: the string has to fulfill outgoing boundary conditions at the horizon, and Neumann ones at the D7 brane. This cannot be done analytically
for generic masses and, by construction, it only yields $\mu$ in the slow $v\to 0$ limit. 
\begin{figure}[ht]
\begin{center} \includegraphics[scale=0.8]{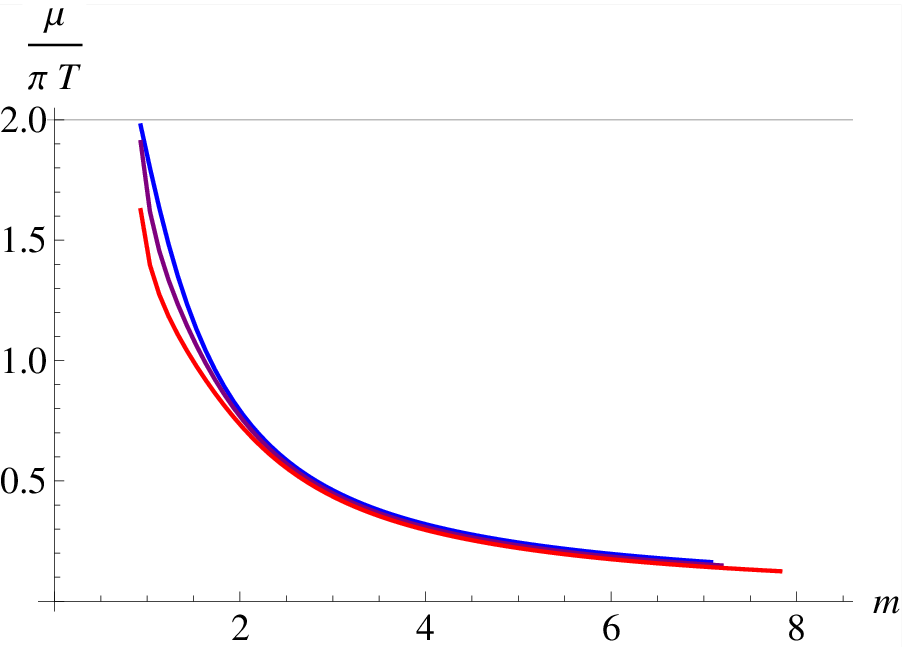}~~ \includegraphics[scale=0.8]{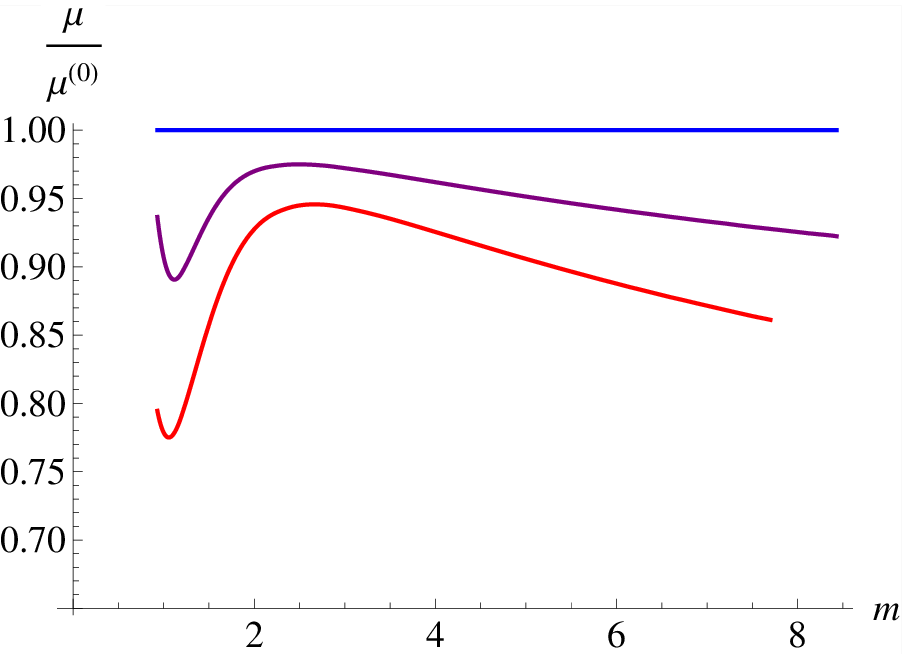} 
\caption{\em{On the left we show the dimensionless quantity $\mu/\pi T$ as a function of the quark bare mass. The plots stop at $m\sim 0.92$ where the
constituent mass vanishes. The horizontal line indicates the
conjectured upper bound for this quantity. The ratio of the drag coefficient in the flavored background with respect to the conformal value is plotted
on the right, as a function of the quark bare mass $m$. The different colors correspond to different values of the parameter $\e_h=0,0.25,0.5$ (from
blue to red).\label{fig:dragcoef}}}
\end{center}
\end{figure}

\begin{figure}[ht]
\begin{center} \includegraphics[scale=0.8]{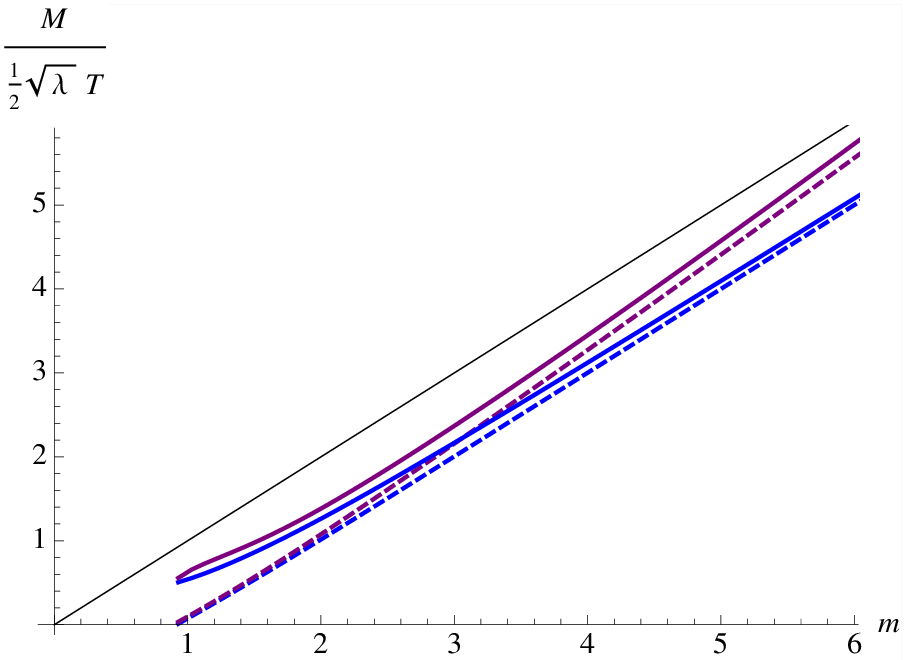}~~ \includegraphics[scale=0.8]{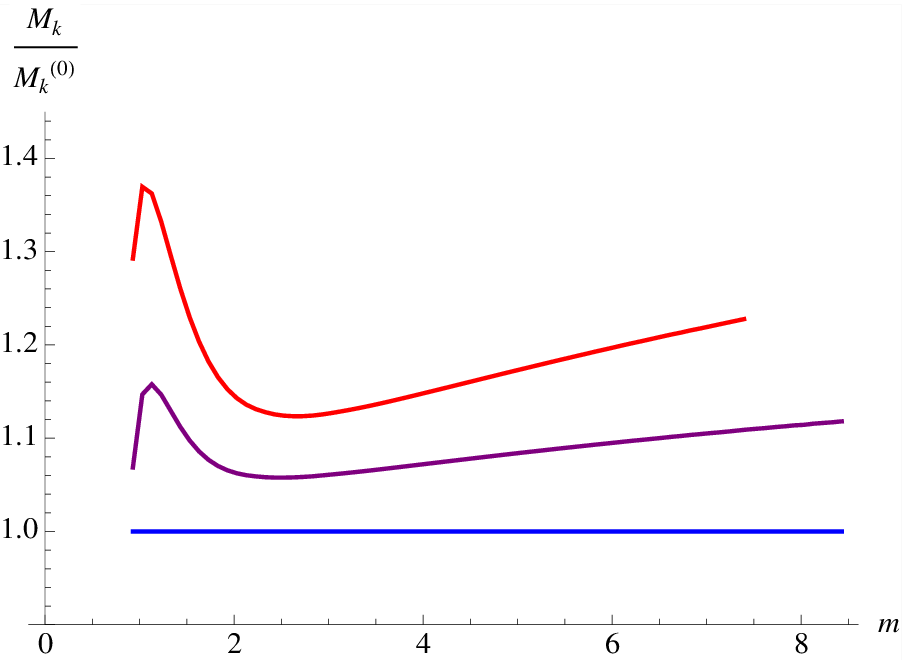} 
\caption{The plot on the left shows the constituent mass $M_c$ (dashed line) and the kinetic mass $M_{kin}$ (plain line) as a function of the bare mass
$m$. On the right we represent the ratio of the kinetic mass in the flavored background with respect to the kinetic mass in the unflavored background,
as a function of $m$. Different colors correspond to values of the parameter $\e_h=$ 0 (blue), 0.25 (purple) and 0.5 (red).
\label{fig:mk}}
\end{center}
\end{figure}

By solving analytically the limit of small masses $M_c\to 0$ ($m\to 0.92$), the authors of 
\cite{herzog1} postulated the existence of an upper bound for the friction coefficient
\be
\mu^{(0)} \leq 2\pi T \ .
\ee

For nonzero flavor density, we observe from figure \ref{fig:dragcoef}, that the friction coefficient is decreased and, in particular, the bound is respected.

Combining the knowledge of $\mu$ with that of $\mu M_{kin}$ previously calculated in \eqref{muMkin} allows to solve for the
kinetic mass. This is, by definition, velocity independent, and should only depend upon the bare mass $m$. The result is shown in the plot \ref{fig:mk}. Both the constituent mass, and the kinetic mass, increase as the flavor density is switched on. The blue curves were obtained in \cite{herzog1} and the continuous (dashed) blue curve represents the kinetic mass $M_{kin}^{(0)}$ (constituent mass $M_c^{(0)}$) at zero flavor density. $M_{kin}^{(0)}$ is a monotonous function of $m$ which has a minimum, for $m\to 0.92$ at $\sqrt{\lambda^{(0)}} T/4$. The purple curves plot the same magnitudes for $\epsilon_h=0.25$.
These curves are only trustable in this interval of masses.

\section{Jet quenching}\label{sec:jetquenching}

The jet quenching parameter measures the momentum broadening of a quark traversing the quark-gluon plasma. If we assume that the heavy quark
undergoes a Langevin diffusion process inside the plasma, the diffusion constant associated to this process is simply related to the jet quenching
parameter. At a classical level, the Langevin diffusion reduces to an ordinary diffusion process. 

The Langevin quantum description implies the presence of a stochastic force acting on the quark, both in the direction longitudinal and transverse to
the motion of the quark. Let us denote the two components of this force by $F_\perp,F_\parl$ respectively. The two-point correlator of this force
defines the diffusion constant $\k$. Namely, at sufficiently long times (with respect to the typical correlation time of the medium, roughly the inverse
temperature), we can write $\langle F(0) F(t) \rangle = \k \delta(t)$, for each longitudinal and transverse direction. 

As it has been shown \cite{gubserjq}, this allows to determine the momentum broadening of the quark in terms of the diffusion constant. We adopt the
definition of the jet quenching parameter as the average momentum squared of the quark divided by the distance the quark has travelled, $\hat q =
\langle p^2 \rangle/L$. Keeping in mind the definition of the diffusion constant, the average momentum squared in the longitudinal and transverse
directions can be obtained in terms of $\k_\parl$ and $\k_\perp$ respectively. This means that, for large enough times $t$, the quark acquires a
transverse momentum that averages to zero but has quantum fluctuations $\langle p_\perp^2 \rangle = 2 \k_\perp t$. The longitudinal momentum has a
classical part related to the drag force and quantum fluctuations from the stochastic force $\langle \Delta p_\parl^2 \rangle = \k_\parl t$.

The diffusion constants can be computed holographically using the prescription for the holographic retarded correlators \cite{sonherzog}. In fact,
they are given by the zero frequency limit of the imaginary part of the retarded correlator, $\k = 2 T_s \lim_{\o\to0} \imm G_R(\o)$. The procedure
implies finding the solution for the trailing string fluctuations with appropriate boundary conditions and evaluate the bulk-to-boundary correlator
that, in turn, yields the retarded correlator. We apply the steps of \cite{gubserjq,casalteaney,iancu}, where the authors obtained the diffusion
constants for the $\cN = 4$ plasma, to the background with backreacted flavors. The aim of this section is to compute the jet quenching parameters in
the flavored background using the following formulae:
\beqa
\hat q_\perp = 2 {\k_\perp \over v}\;, \quad \hat q_\parl = {\k_\parl \over v}\;.
\eeqa

In order to compute the retarded correlator corresponding from the trailing string, the string ansatz must include the quantum fluctuations $\del
X^i$
\beqa\label{ansatzfl}
X^1 = vt + x(r) + \del X^1(r,t),\quad X^a = \del X^a(r,t), \quad a=2,3\;.
\eeqa
By a change in the time coordinate $\tilde t = t+\z(r)$, with $\z'(r) = {g_{tr} / g_{tt}}$ \cite{Gursoy:2010aa}, we obtain a quadratic action for the fluctuations with no
mixed $\partial_t \del X \,\partial_r \del X$ terms
\beqa\label{NG2}
S_{NG}^{(2)} = -{1\over 2\pi\a'} \int dt\,dr\, \sum_{i=1}^3
\left[ \cG_{i}^{rr} (\partial_r \del X^i)^2 + \cG_{i}^{tt} (\partial_t \del X^i)^2 \ri] \ ,
\eeqa
with
\beqa\label{Gab}
\cG_{\perp}^{\a\b}
= Z^{-2} \; \cG_{\parl}^{\a\b}
= {1 \over 2} \le( \begin{array}{cc} -G_{xx} H^{-1} & 0 \\ 0 & H \end{array} \ri) \;,
\eeqa
and
\beqa
H &=& e^{\phi \over 2} {\sqrt{G_{xx} \le(G_{tt} G_{xx} - e^{\phi_s - \phi} G_{tt,s} G_{xx,s} \ri) \le( G_{tt,s} G_{xx} - G_{tt} G_{xx,s} \ri) \over G_{tt}
G_{rr} G_{xx,s}}} \label{hache} \ , \\
Z &=& e^{(\phi_s-\phi) \over 4} {G_{xx,s}\over G_{xx}} { \sqrt{ G_{tt} G_{xx} - e^{\phi_s-\phi} G_{tt,s} G_{xx,s} \over G_{tt,s} G_{xx} - G_{tt} G_{xx,s}}}.
\label{zeta}
\eeqa
Here the subscript $s$ indicates functions evaluated at the worldsheet horizon $r_s$.
The retarded correlator is then obtained as
\beqa\label{GR}
G_R(\o) = -\le[\Psi_R^*(r,\o) \cG^{rr} \partial_r \Psi_R(r,\o)\ri]_{\rm boundary} \ ,
\eeqa
where $\Psi_R(r,\o)$ is the retarded solution to the fluctuation equation, $\Psi(r,\o) = e^{i \o \tilde t} \del X(r,\tilde t)$, with boundary conditions
given by $\Psi_R(r,\o)\to 1$, at the boundary, and $\Psi_R(r,\o) \approx (r-r_s)^{i \o / 4 \pi T_s}$, at the worldsheet horizon. The fact that the
fluctuations behave like infalling waves with temperature $T_s$ at the worldsheet horizon implies that $T_s$ appears in the aforementioned Einstein
relation.

In fixing the
boundary normalization condition, we assume that the fluctuation has a regular expansion in the UV. By this we mean that for large radii the leading
behavior of $\Psi$ is constant. Indeed, using the expressions \eqref{bulkmettau}-\eqref{solphi} and expanding for a large enough $r$ such that $r_s
\ll r \ll r_h e^{1/\e_h}$, we arrive to the expected expansion for the wave function: $\Psi \simeq C_{\rm source} + C_{\rm vev} e^{-\phi/2}/r^3$.
The normalization condition can be set at a cutoff radius $r_{min}$ (related to the probe quark mass $M_c$ by equation \eqref{constmass}). This avoids the
problem of going beyond the truncated expansion in $\e_h$ for the bulk fields and we can stay in the regime where $r_s \ll r_{min} \ll r_h
e^{1/\e_h}$. Moreover, the zero frequency limit of $\imm G_R$ is independent of this cutoff. Hence we can consistently derive the diffusion
constants (and consequently the jet quenching parameters) with a cutoff and arrive to a final cutoff-independent result.

In order to compute the zero frequency limit of the imaginary part of the retarded correlator we do not need the full solution, but only its
asymptotics. Moreover, since $\imm G_R$ is a conserved flux, it can be evaluated at an arbitrary point in the radial coordinate (not necessarily at the
boundary). In particular, it is easier to evaluate it at the horizon, using the in-falling wave function expression $\Psi_R(r) \simeq \Psi_s
(r-r_s)^{i \o / 4 \pi T_s}$. Substituting this asymptotic into the formula \eqref{GR} for the retarded correlator and reminding the definition of the
diffusion constant, we obtain
\beqa
\k = {1\over \pi \a'} B_s^2 T_s |\Psi_s|^2,
\eeqa
with $B_s$ defined by
\beqa
\cG^{rr}(r) \simeq 4 \pi T_s B_s^2 (r-r_s) + \cO(r-r_s)^2, \qquad r\to r_s.
\eeqa
The constant $\Psi_s$ is evaluated by matching the low frequency limit of the retarded solution with the exact zero frequency solution and imposing
the normalization condition. Summarizing the results of this computation, we get $\Psi_s = 1$.
Moreover, deriving the asymptotics of equation \eqref{Gab}, we obtain $B_{\perp}^2 = e^{\phi/2} G_{xx}$ and similarly for $B_\parl^2 = e^{\phi/2}
G_{xx} Z^2$. Therefore, the diffusion constants and jet quenching parameters read
\beqa
\hat q_\perp 
&=& ~ 2\pi T^3 \sqrt{\lambda^{(0)}} \frac{ \sqrt{\gamma}}{v} \left( \rule{0mm}{6mm} 1+\frac{ \epsilon_h}{8}(1 + v^2 + 2\log\gamma) + \right.
\\
&&\left. + \frac{\epsilon_h^2}{384} (15 + 14 v^2 - 3v^4+4(7+3v^2)\log\gamma + 36(\log\gamma)^2 + 12\log_2(v^2)) \right)
 \ , \nonumber \\
\hat q_\parl 
&=& ~ 2\pi T^3 \sqrt{\lambda^{(0)}} \frac{ \sqrt{\gamma}}{v} \left( \rule{0mm}{6mm}1+\frac{ \epsilon_h}{8}(1 + 3v^2 + 2\log\gamma) + \right.
\\
&&\left. + \frac{\epsilon_h^2}{384} (15 + 42 v^2 +9v^4+4(19+9v^2)\log\gamma + 36(\log\gamma)^2 + 12\log_2(v^2)) \right) \ . \nonumber
\eeqa
Here we used the asymptotic expansion of the background metric component and the relation between $r_s$ and $T$. We remind that this expansion is valid
as long as the velocity satisfies $\e_h \log \g \ll 1$.

\begin{figure}[ht]
\begin{center}
\begin{tabular}{cc}
\includegraphics[scale=0.75]{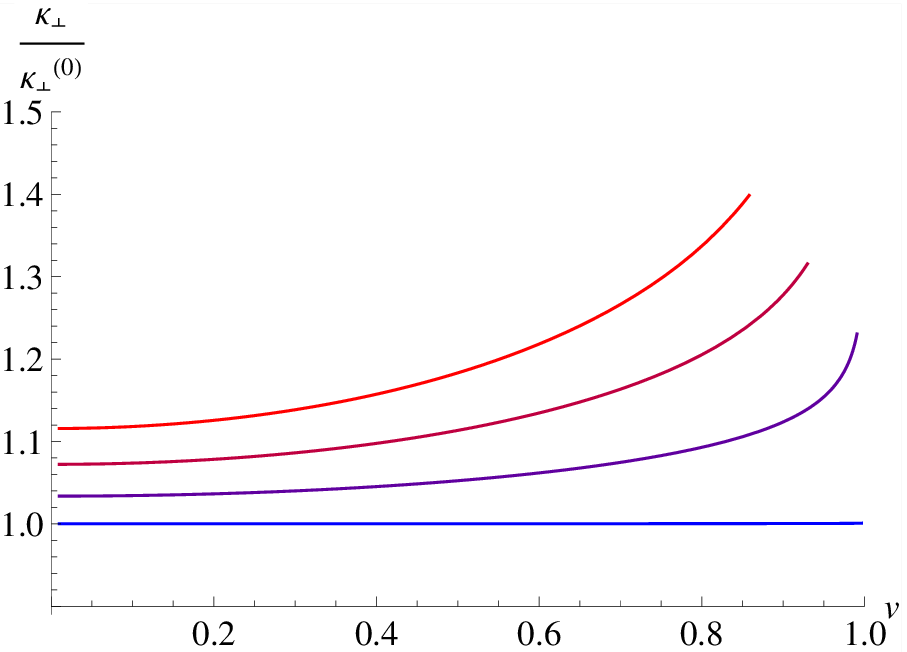} ~~&~~ \includegraphics[scale=0.75]{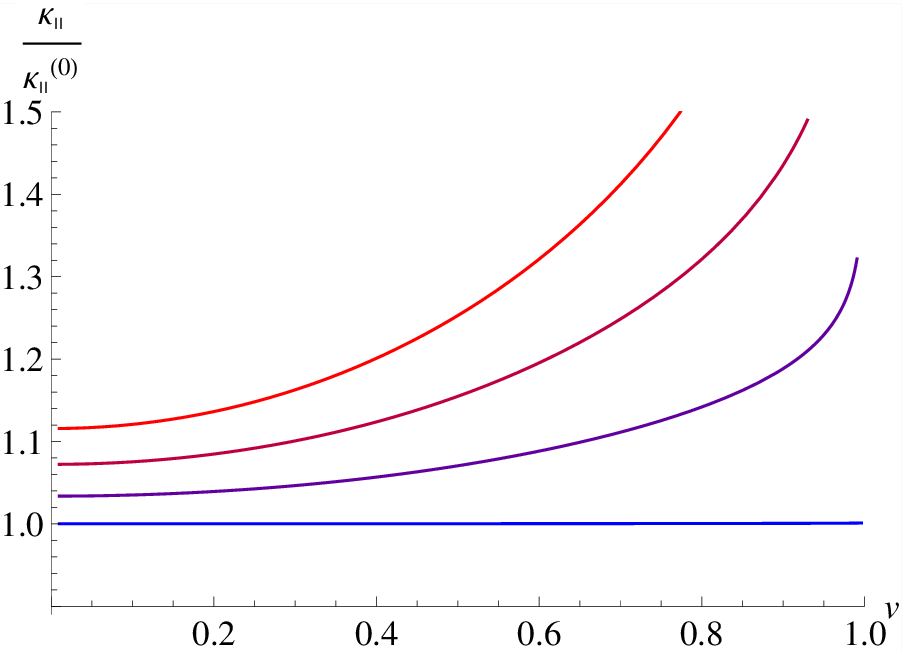}
\end{tabular}
\end{center}
\caption{\label{fig:kappa} \em{The ratio of the diffusion constants $\kappa_\perp,\kappa_\parl$ to their value in the unflavored background are shown as a function of the
velocity $v$, for different values of the perturbative parameter $\e_h=0,0.25,0.5,0.75$ (from bottom up, or from blue to red). This ratio is as well the ratio of the jet quenching
parameters with respect to their supersymmetric values, since they are proportional to the diffusion constants.}}
\end{figure}
\begin{figure}
\begin{center}
\begin{tabular}{cc}
\includegraphics[height=4.7cm]{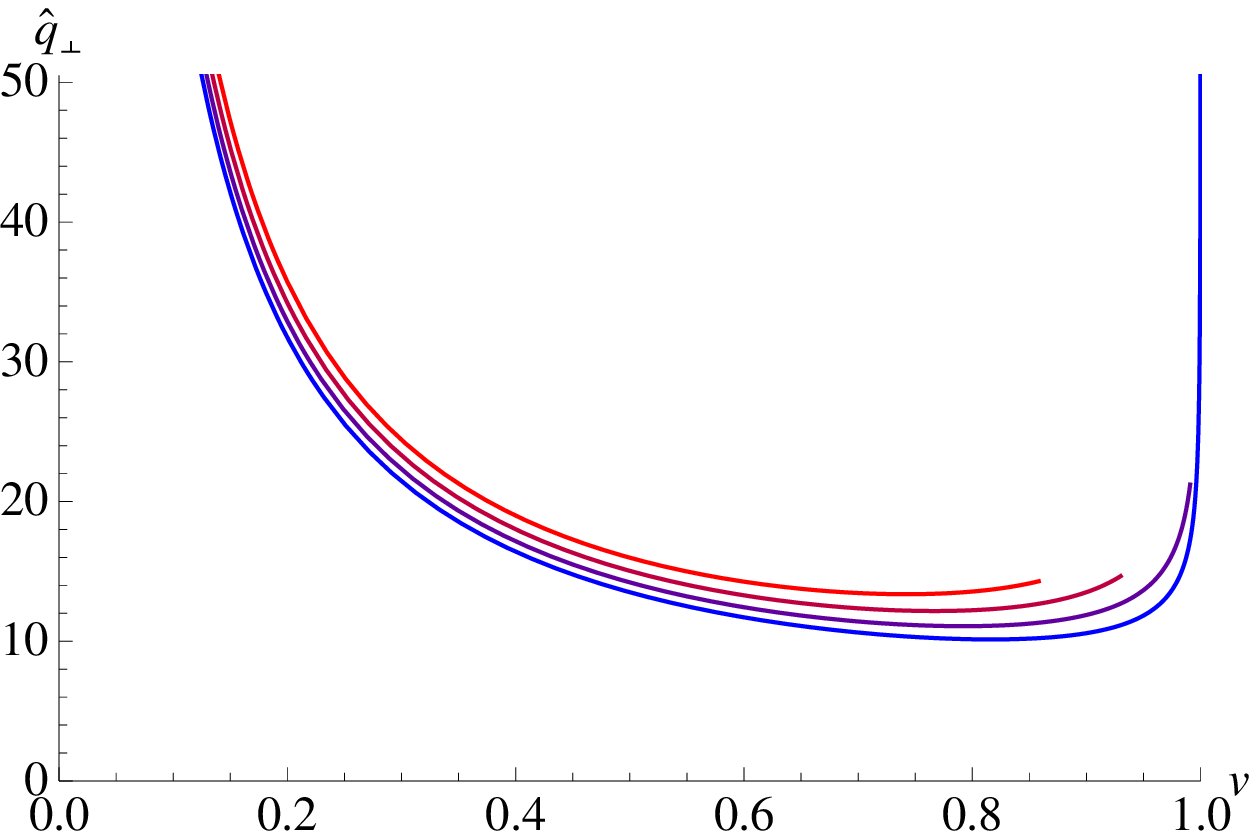}~~ & ~~ \includegraphics[height=4.7cm]{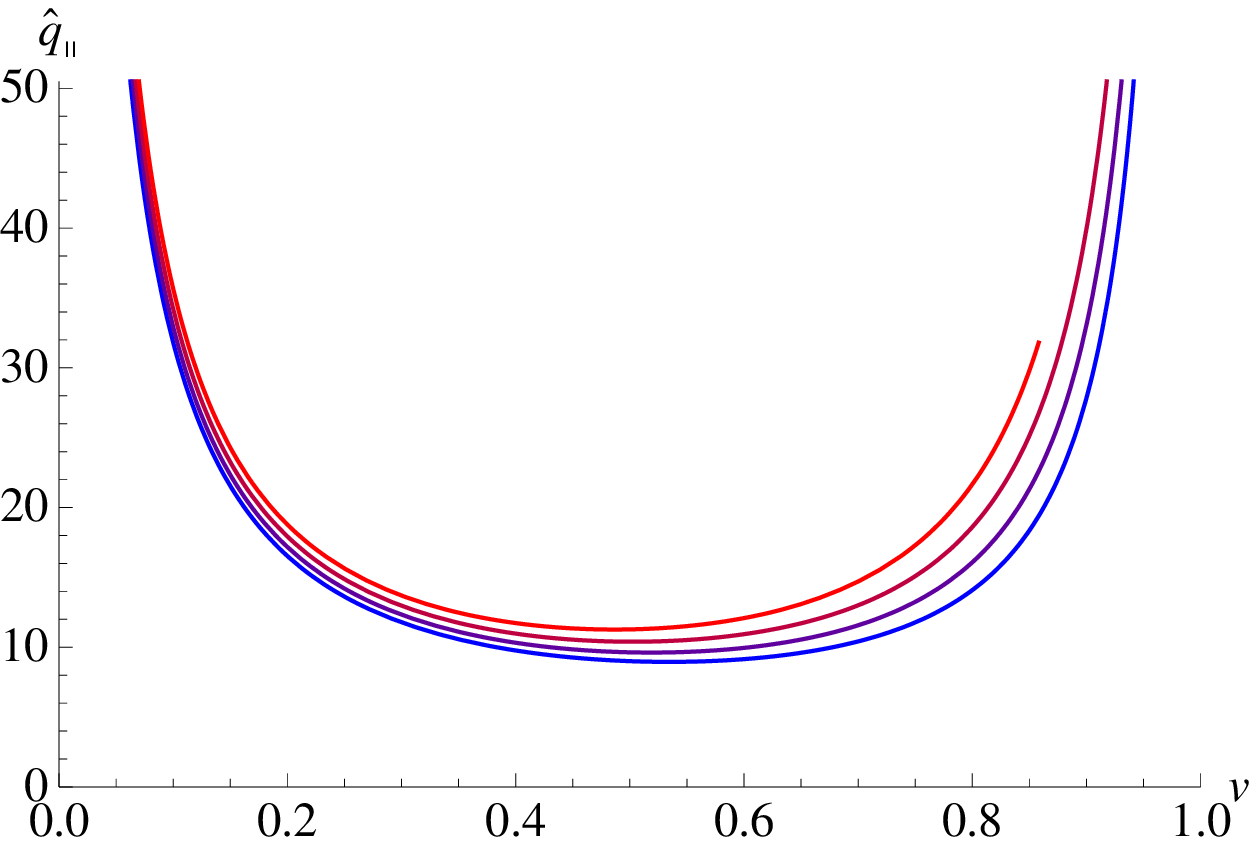}
\end{tabular}
\end{center}
\caption{\label{fig:q} \em{Here, the jet quenching parameters $\hat{q}_\perp,\hat q_\parl$ (in units of $\sqrt{\lambda^{(0)}}T^3$) are shown as a function of the velocity
$v$, at a fixed temperature, and for different values of the perturbative parameter $\e_h=0,0.25,0.5,0.75$.}}
\end{figure}

The corrections coming from the flavor backreaction in $\hat q_\perp$ are the same as for the drag force with the addition of corrections coming from
the worldsheet temperature in equation \eqref{TsT}. We conclude that the jet quenching parameter related to the transverse momentum broadening in the
flavored background is increased with respect to the value it takes in the unflavored background. The same conclusion applies to the momentum acquired
in the longitudinal direction, since also for $\hat q_\parl$ the corrections are always positive. Visually, this is shown in the plots of figure
\ref{fig:kappa}, where the ratio of the diffusion constants (equivalently the ratio of the jet quenching parameters) is represented.

In figure \ref{fig:q} the jet quenching parameters are plotted, relative to the transverse directions and the longitudinal direction respectively.

The Einstein relation in terms of the jet quenching parameter and the drag force can be written as $$\hat q = {4 T_s \over v^2} f,$$ where we have to
use the
worldsheet temperature. This relation is well known to be valid in the $\cN =4$ supergravity dual, as well as in dilaton-gravity models such as
\cite{Gursoy:2010aa}, and remains unchanged in the backreacted background, order by order in $\e_h$.

\section{Conclusions}
The physics of quarks propagating through dense media is a subject of intense research nowadays. The fact that this problem, most probably, involves
QCD at strong coupling implies a challenge to the traditional perturbative methods. Holography provides a valuable tool, in spite of the fact that the
gravity dual model is only qualitatively related to quantum chromodynamics. The traditional objections to the presence of supersymmetry are bypassed
at finite temperature, where supersymmetry is indeed broken. In the present work we have analyzed the properties of probes on a background that was
constructed some time ago \cite{D3D7QGP}. This background is analytic. However, the price to pay is that it is a perturbative deformation in powers
of a parameter which is roughly the Veneziano parameter $\sim N_f/N_c$. The use of this dual geometry only makes sense in the spirit of effective
field theory. The background comes with a cutoff $ r_*$, beyond which the predictions blow up because of the presence of a Landau pole. Conversely all
the physical predictions at scales much lower than the cutoff $r\ll r_*$ are robust and unique up to corrections which die off with powers of $r/r_*$.
In the present context, the lowest cutoff energy that one can consider is given by the plasma temperature (or the horizon radius $r\geq r_h$). When
examining the physics in the furthest possible IR, it makes sense to discard the UV corrections, or equivalently send $r_*\to \infty$, and work with
the effective metric written in \eqref{solttxx}-\eqref{solphi}. To make sense of the analytic expansion in terms of the parameter $\e_h$, related to
the Veneziano factor, the effective energy bound is $r<r_\Lambda\sim r_h e^{1/\epsilon_h}$. This implies that the mass of the quark probes cannot be
taken to be very large, namely the relevant range for masses is $M_q \lesssim 3.5$ GeV.

 Given all this, and moreover, that among all possible ``conifold like" models we have chosen the simplest one, namely $AdS_5\times S_5$, we shall
only trust qualitatively our results. At a practical level this amounts to finding the sign of the first order correction to physically motivated
quantities. Whether it is positive, negative or vanishing is all we can say within the present effective approach. Of course, as emphasized in the
text, there is not a unique prescription as to what two theories should be compared. The choice both here and in \cite{D3D7QGP}, \cite{D3D7QGPb} has
been to keep the total amount of degrees of freedom constant, as measured by the energy density. Using the entropy density would have given
qualitatively equivalent results. The phenomenological outcome of this work can be summarized in few sentences. The effective thermal masses (constituent $M_c$ and kinetic $M_{kin}$) increase in a flavored theory as compared with their values in an unflavored one. The meson melting transition occurs at a lower value of the ratio $M_q/T$. For fixed quark mass this implies a higher temperature. The $q\bar q$ screening length is also diminished. The conductivity is reduced, which is consistent with the fact that the drag force increases. The flavor diffusion constant is lowered, but the broadening of jets, as parametrized by the jet quenching parameter, is enhanced. All in all, the consistent picture that arises seems to point out in the direction of saying that the fundamental degrees of freedom, have a larger cross section than the adjoint ones.

\section*{Acknowledgments}

We would like to thank Aldo Cotrone, Elias Kiritsis, Nick Evans, Carlos N\'u\~nez, Francesco Nitti, Ioannis Papadimitriou, Alfonso V. Ramallo and Balt van Rees, for useful conversations and comments.

J.T.'s research is supported by the Netherlands Organization for Scientic Research (NWO) under the FOM Foundation
research program.
J.M. and L.M. are supported by the MICINN and FEDER (grant
FPA2008-01838), the Spanish Consolider-Ingenio 2010 Programme CPAN
(CSD2007-00042), and the Xunta de Galicia (Conselleria de Educaci\'on
and grant INCITE09-206-121-PR). 

\appendix

\section{Derivation of the conductivity}\label{app:conductivity}

We will mimic \cite{arXiv:0705.3870} to obtain the conductivity for the probe D7-brane. To simplify notation let us write for the pullback metric
\be\label{pullback2}
\dd s_8^2 = g_{tt}(r) \dd t^2 + g_{xx}(r) \dd x^i \dd x^i + g_{rr}(r,\chi(r)) \dd r^2 + ds^2_{\hbox{\tiny int}}(\chi(r),\theta^a) \ ,
 \ee
 where $\theta^a= (\theta, \xi,\varphi)$ represent a 3-sphere wrapped by the probe brane, and the pullback metric components are related to the background 10d components given by
 \beqa
 g_{tt} &=& G_{tt} \label{bulkgtt} \ ,
 \quad g_{xx} = G_{xx} \ ,
 \quad g_{rr} = G_{rr} + \frac{\gke}{2} \frac{\psi'^2}{1-\psi^2} \ , \\
 ds^2_{\hbox{\tiny int}} &=& \frac{\gke}{4} (1-\psi^2) \left[(w^1)^2\,+\,(w^2)^2\,+\,(w^3)^2\right]+ \frac{G_{\tau\tau}-\gke}{4} (1-\psi^2)^2(\omega^3)^2 \ .
 \eeqa
We have field strength $F=\dd A$ where, as explained in the main text, the gauge field is given by $A= A_0(r)\dd t + (E_x t + A_x(r)) \dd x$. 

The action for the probe brane has the following form in Einstein frame
\be
S = S_{DBI} + S_{CS} = - T_7 \int d^8\chi\, e^{\Phi}\sqrt{-\det( g_8 + e^{-\Phi/2}\hat F)} + S_{CS} \ ,
\ee
where $ds_8^2= ds_E^2$ is the induced line element in the Einstein frame, and $\hat F \equiv 2\pi \alpha' F$. When evaluating the determinant in the DBI action, the compact space will factorize since $F$ does not take values along these directions, the contribution reads (with $a,b=\theta, \psi,\varphi$)
 \be
 \det g_{ab} = \gke^3 (1-\psi^2)^3 \left( 1 + \left( \frac{G_{\tau\tau}}{\gke}-1\right) (1-\psi^2) \right)\, \frac{\sin^2\theta}{64}\equiv
 g_{int}(r,\psi(r))\frac{\sin^2\theta}{64} \ .
\label{gint}
\ee
In this expression we have separated the radial dependence, $g_{int}(r,\psi(r))$, from the angular part. Altogether
\be
\det (g_8+e^{-\Phi/2}\hat F) = e^{-\Phi} \left( g_{rr} (e^{\Phi} |g_{tt}|g^3_{xx}- g_{xx}^2\hat E_x^2 ) + |g_{tt}| g^2_{xx} \hat A_x'^2 - g^3_{xx} \hat A_0'^2\right) g_{int} \frac{\sin^2\theta}{64} \ ,
\ee
hence
 \beqa
S_{DBI}= - V_3 \tilde {\cal N} \int dt dr \, e^{\Phi} \sqrt{ \left( g_{rr} |g_{tt}|g_{xx}-e^{-\Phi}\left(g_{rr} \hat E_x^2 - |g_{tt}| \hat A_x'^2 + g_{xx} \hat A_0'^2\right)\right) g_{xx}^2g_{int} } \ ,
 \eeqa
 where $\tilde {\cal N}=2\pi^2 N_f' T_7$, $V_3=\int d^3x$ and we have used the factorization of the angular variables to obtain the volume of the 3-sphere, $\int d^3\theta \sin\theta/8=2\pi^2$. Notice the length dimensions $[\tilde {\cal N}]=-8$ and $[g_{int}]=6$. 
 
There are first integrals of motion for $A_{0,x}$, which are holographically related to the charge and current densities
\be
 \frac{\delta {\cal L}_{DBI}}{\delta A_0'} = n_q \ , \quad \frac{\delta {\cal L}_{DBI}}{\delta A_x'} = \langle J_x \rangle \ ,
\ee
with dimensions $[n_q] = [j_x] = -3$, since $[{\cal L}_{DBI}] = -5$. Dividing by $2\pi\alpha'\tilde{\cal N}$ and defining the quantities $\tilde n_q = n_q/(2\pi\alpha'\tilde{\cal N})$ and $\langle \tilde J_x \rangle = \langle J_x\rangle /(2\pi\alpha'\tilde{\cal N})$, we can express the radial derivatives of $A_{0,x}$ as
\beqa
\hat A_0' & = &- \sqrt{ \frac{ |g_{tt}| g_{rr} \left(e^{-\Phi} \hat E_x^2- |g_{tt}| g_{xx} \right) \tilde n_q^2 }{g_{xx}\left( |g_{tt}| g_{xx}^3 g_{int}+e^{-\Phi}( |g_{tt}| \tilde n_q^2 - g_{xx} \langle \tilde J_x \rangle)\right) }} \ ,
\label{Normalizableexp1}\\
\hat A_x' & = & - \frac{g_{xx}}{|g_{tt}|}\hat A_0' \ .
\label{Normalizableexp2}
\eeqa

One then defines a special locus (which we name singular shell) as the radius $r_\s$ such that the numerator in the square root of $\hat A_{0}$ vanishes, and fixes the value of $\langle \tilde J_x \rangle$ to make the denominator vanish at the same point, to keep the gauge field real (and, therefore, the Legendre transformed action w.r.t $A_{0,x}$ stays real too)
\beqa
\left. e^{-\Phi} \hat E_x^2- |g_{tt}| g_{xx} ~\right\vert_{r=r_\s} &=& 0 \label{singsheleq} \ ,\\
\left.|g_{tt}| g_{xx}^3 g_{int}+e^{-\Phi}( |g_{tt}| \tilde n_q^2 - g_{xx} \langle \tilde J_x \rangle^2)~\right\vert_{r=r_\s} &=& 0 \ .\label{alskdjl}
\eeqa
Equation (\ref{singsheleq}) defines implicitely the sigular shell $r_\s(r_h)$.
Notice that for a metric in the form given in \eqref{bulkgtt} we have, from equation (\ref{singsheleq}), that the singular shell and the horizon radius are related by
\be
r_\s^4 = r_h^4 + R^4 e^{-\Phi}\hat E_x^2 \ .
\ee
From equations (\ref{singsheleq} -\ref{alskdjl}), and using the definitions given in this appendix, we obtain the expression
\be
\langle J_x \rangle = \sqrt{(2\pi\alpha' \tilde {\cal N})^2g_{xx} g_{int} + e^{-\Phi} n_q^2 g_{xx}^{-2}}\Bigg|_{r_\s} 2\pi\alpha' E_x\, . \label{current2}
\ee
Which gives the following expression in terms of the bulk components:
\beqa
\sigma_{DC} &=& \frac{\langle \tilde J_x \rangle}{E_x} \\&=& \sqrt{(2\pi\alpha')^4 \tilde {\cal N}^2 G_{xx} \gke^3(1-\psi^2)^3\left( 1 + \left( \frac{G_{\tau\tau}}{\gke}-1\right)\tilde\psi^2\right) 
 + (2\pi\alpha')^2 e^{-\Phi} n_q^2G_{xx}^{-2}} ~ \Bigg|_{r=r_\s}.\non
\eeqa

Taking the limit $E_x\to0$ is equivalent to taking the limit $r_\s\to r_h$, where 
\be
(2\pi\alpha')^4 \tilde {\cal N}^2 G_{xx} \gke^3(1-\psi^2)^3 = \frac{N'^2_f (N_c^{(0)})^2}{2^{4}\pi^{2}}T^2 \left(1 - \frac{\epsilon_h}{6}\right) (1-\psi_h^2)^3 \ ,
\ee
and also
\be
 (2\pi\alpha')^2 e^{-\Phi} n_q^2G_{xx}^{-2} = 4 e^{-\phi}\frac{ n_q^2}{\lambda_h^{(0)} \pi^2 T^4}\left( 1 - \frac{\epsilon_h}{4}\right) \ ,
\ee
which gives the result quoted in (\ref{conductivityresult}). As a final remark concerning dimensions, notice that for the conductivity, $[\sigma_x]=-1$, is consistent with Ohm's law, since $[j^\mu]=-3$ whereas $[E] = [F_{\mu\nu}] = -2$.

 Following again the procedure stablished by Karch and O'Bannon in \cite{arXiv:0705.3870} we can study the drag force from the conductivity. 
For very heavy carriers we expect a classical equation of forced motion
\be
\frac{d p}{dt} = -\mu p + f
\ee
In the stationary situation, with $f=E_x$, this will give rise to a steady current
\be
j_x = \sigma_x E_x = n_q v, \label{currentv}
\ee
with $n_q$ the density of charge carriers and $v$ their velocity. In the very massive probe limit we can discard the first term in the expression (\ref{current2}) and compare it with(\ref{currentv}), obtaining
\be
v = \frac{(2\pi\alpha') E_x}{e^{\Phi/2} g_{xx}}\Bigg|_{r_\sigma}.\label{dkdajll}
\ee
Using this result and the definition of the singular shell \eqref{singsheleq}, as well as the explicit values of the induced metric components in terms of the bulk metric components
 we find that the velocity and drag force are given by
\be
v^2 =\left. \frac{|G_{tt}|}{G_{xx}}\right\vert_{r=r_\sigma} "~~~~~;~~~~f = \left. \frac{1}{2\pi\alpha'} e^{\Phi/2} \sqrt{G_{xx}|G_{tt}| } \right\vert_{r_\sigma} \, . \label{aslkxasdl} 
\ee
 
\section{Drag Force and Jet Quenching}
Central to the discussion of all these effects is the action for a fundamental string
\begin{equation}\label{NG}
S_{NG} = -\frac{1}{2\pi \a'}\int d^2\sigma \sqrt{-\det g_{\alpha\beta}} \;, \qquad g_{\alpha\beta}= e^{\Phi/2} G_{MN} \partial_{\alpha}X^{M}
\partial_{\beta}X^{N}.
\end{equation}
where the string worldsheet coordinates in the static gauge $\s^{0}=t,\quad \s^1=r$ are labelled by $\a,\b=0,1$ and $G_{MN}$ is the bulk metric in the
Einstein frame defined in equation \eqref{deformedads5bh}, with components given by equation \eqref{bulkmettau}. 
 From here the momentum currents
can be readily derived
\be
\left(\begin{array}{c}\pi^0_x \\ \pi^1_x \end{array} \right) =
\frac{e^{\Phi/2}}{2\pi\alpha' \sqrt{-g}}
\left(\begin{array}{c}\displaystyle \Grr\Gxx\dot x \\ \rule{0mm}{6mm}
\displaystyle \Gtt\Gxx x' \end{array} \right)\label{momenta}
\ee

Let us consider first the stationary solution:
the ansatz for the classical trailing string reads
\beqa
X^1=vt+x(r),\quad X^i=0, \quad i=2,3\;,\label{trstran}
\eeqa
where $v$ is the velocity of the quark moving along the $X^1$ direction and $x(r)$ determines its (steady state) position. 
The classical on-shell action for the trailing string only depends on the radial derivative of $x$. Therefore, the momentum conjugate to $x$,
$\pi^1_x$, is constant along the radial direction. Moreover, $\pi^1_x$ is related to the momentum loss of the quark, due to the friction of the medium as
the quark is moving in the $X^1$ direction, $\pi^1_x = - dp_1/dt$. This is the drag force that we want to compute.

To proceed, we solve the equation for $x'$ in terms of the constant $\pi_x$ and of the background metric. This yields
\beqa\label{xprime sol}
x' = 2 \pi \a' \pi_x \sqrt{ {G_{rr} \over G_{tt} G_{xx}} {G_{tt} + G_{xx} v^2 \over (2 \pi \a' \pi^1_x)^2 + e^\Phi G_{tt} G_{xx}} }\;.
\eeqa
Reality of this expression is ensured if the denominator and the numerator both change sign simultaneously at the same point $r_s$
defined by 
\be 
 v^2= {|G_{tt}| \over G_{xx}}\Bigg|_{r_s} ~~~~~~,~~~~~
{dp_1 \over dt} = - \pi^1_x = - {1 \over 2\pi \a'} e^{\Phi/2} G_{xx} v \bigg|_{r_s}.
\label{–lkda–sdlk–}
\ee 
From the first equation and \eqref{solttxx} we can solve for $r_s$ in terms of $r_h$ as
$r_s =\sqrt{\gamma} \, r_h\, . \label{rdese}$ with $\gamma = (1-v/2)^{-1/2}$.
Also, from the second, we have the following implicit expression for the drag force
\be 
f=-\frac{dp_1}{dt} = {1 \over 2\pi \a'} e^{\Phi/2} G_{xx} v \bigg|_{r_s} \label{sd–flfk–lkf–}
\ee
Comparing \eqref{–lkda–sdlk–} and \eqref{sd–flfk–lkf–} with \eqref{aslkxasdl}, we find consistency provided $r_\sigma = r_s$.

In order to proceed with the fluctuations it is most convenient to go to a different gauge.
\be
\tilde \sigma_a = \left(\rule{0mm}{4mm} \tilde \sigma_0 = t + \zeta(r),\tilde \sigma_1 = r \right)~~~;~~~ \zeta'(r) = \frac{g_{01}}{g_{00}}= \frac{
\dot x x' G_{xx}}{G_{tt} + \dot x^2 G_{xx}},
\ee
for which the induced metric is diagonal for the same trailing string ansatz as \eqref{trstran}
\be
\tilde g_{ab} = \left(
\begin{array}{cc}
 \Gtt +\dot x^2 \Gxx & 0 \\
0 & \Grr + x'^2 \Gxx - \displaystyle\frac{(\dot x x' \Gxx)^2}{G_{tt} + \dot x^2 G_{xx}}
\end{array}
\right).
\ee
One sees that $r_s$ is the location of an event horizon of the induced worldsheet metric $\tilde g_{00}(r_s)=0$, for which, performing the standard manipulations one can compute the associated Hawking temperature 
\beqa
\label{TsT}
T_s &=& \frac1{\b_s} = {1 \over 4\pi} \sqrt{ \le( G_{tt}' + G_{tt} {\Phi' \over 2} \ri)^2 G_{xx}^2 - G_{tt}^2 \le( G_{xx}'' + G_{tt} {\Phi' \over 2}
\ri)^2 \over G_{rr} G_{tt} G_{xx}^2} \Bigg|_{r_s} \, ,\label{genTs}
\eeqa
from which the result given in \eqref{wstemp} follows upon expansion up to second order in $\e_h$
This, as usual, defines the worldsheet temperature $T_s$ in terms of the bulk temperature $T$. 

In order to compute the jet quenching parameter, we have to resort to the analysis of fluctuations.
The ansatz is 
\be
X^1 = vt + x(r) + \delta X^1(t,r)~~~;~~~X^i = \delta X^i ~~~i=2,3.
\ee
Expanding \eqref{NG} to second order we have (in $\tilde \sigma_a $ coordinates)
\be
S^{(2)}_{NG} = -\frac{1}{2\pi\alpha'}\int d^2\sigma~
\left(\tilde {\cal G}^{ab}_{\|} \delta X^1_{,a}\delta X^{1}_{,b} +\sum_{i=1}^2 \tilde {\cal G}^{ab}_{\perp} \delta X^i_{,a}\delta X^{i}_{,b} 
\right),
\label{diagsecordact}
\ee
 where the (diagonal) tensors $\tilde {\cal G}^{ab}_{\|\, ,\perp}$ are given in the main text \eqref{hache} \eqref{zeta}.
 Now we can derive the equations of motion in this gauge. Let $\partial_i = \partial_{\tilde \sigma^i}$
\beqa
\partial_r\left( e^{\Phi/2}{\cal G}^{11}_I \partial_r \Psi_I \right) - \omega^2 e^{\Phi/2}{\cal G}^{00}_I \Psi_I = 0~~~~~I = \perp, \|
\eeqa
with
\beqa
\left\{
\begin{array}{rcl}
X^{2,3}(\tilde\sigma^0, \tilde \sigma^1=r) &=& e^{i\omega \tilde \sigma^0} \Psi_\perp(\omega,r ) \\
X^1(\tilde \sigma^0,\tilde \sigma^1=r) &=& e^{i\omega \tilde \sigma^0} \Psi_\|(\omega, r) 
\end{array}
\right.,
\eeqa
or, explicitly,
\beqa
\partial_r\left( H \partial_r \Psi_{\perp}\right) + \omega^2 \frac{G_{xx}}{H}\Psi_{\perp} &=& 0 \nonumber\\
\partial_r\left( Z H \partial_r \Psi_\|\right) + \omega^2 \frac{G_{xx} Z}{H} \Psi_\| &=& 0\, ,
\eeqa
were $H$ and $Z$ were given in the main text \eqref{hache} and \eqref{zeta}.


\end{document}